\documentclass[usenatbib]{mnras}

\usepackage{mathptmx}
\usepackage[T1]{fontenc}
\usepackage[british]{babel}

\usepackage[utf8]{inputenc}
\usepackage{lmodern}

\usepackage{ae,aecompl}

\usepackage{graphicx}	
\usepackage{amsmath}	
\usepackage{amssymb}	

\usepackage{tablefootnote}
\usepackage{savefnmark}
\usepackage{epstopdf}
\usepackage{natbib}
\usepackage{hyperref}
\usepackage{caption}

\usepackage{pdfpages}

\title[WASP-92, WASP-93, WASP-118: TTV and stability]{WASP-92, WASP-93 and WASP-118: Transit timing variations and long-term stability of the systems}

\author[Gajdo\v{s} et al.]{Pavol Gajdo\v{s}$^{1}$\thanks{E-mail: pavol.gajdos@student.upjs.sk},
Martin Va\v{n}ko$^{2}$, 
Mari\'{a}n Jakub\'{i}k$^{2}$, 
Phil Evans$^{3}$,
Marc Bretton$^{4}$,\newauthor
David Molina$^{5}$, 
St\'{e}phane Ferratfiat$^{6}$, 
Eric Girardin$^{7}$,
Sn\ae{}varr Gu\dh{}mundsson$^{8}$,\newauthor
Francesco Scaggiante$^{9}$,  
\v{S}tefan Parimucha$^{1}$
\\
$^{1}$Institute of Physics, Faculty of Science, Pavol Jozef \v{S}af\'arik University, Ko\v{s}ice, Slovakia\\
$^{2}$Astronomical Institute, Slovak Academy of Sciences, 059 60 Tatransk\'a Lomnica, Slovakia\\
$^{3}$El Sauce Observatory, Chile\\
$^{4}$Baronnies Proven\c{c}ales Observatory UAI B10, France\\
$^{5}$Anunaki Observatory, Madrid, Spain\\
$^{6}$Astronomes  Amateurs  Aixois  de  l'Observatoire  de  Vauvenargues, Vauvenargues, France\\
$^{7}$Grand-Pra private observatory, Switzerland\\
$^{8}$Nes Observatory, Iceland\\
$^{9}$Gruppo Astrofili Salese Galileo Galilei S.M di Sala, Venezia, Italy\\
}

\date{Accepted 2019 March 7. Received 2019 March 6; in original form 2018 November 6}

\pubyear{2019}

\begin{document}
\setcounter{page}{3580}
\label{firstpage}
\volume{485}
\pagerange{\pageref{firstpage}--\pageref{lastpage}}
\maketitle

\begin{abstract}
We studied three exoplanetary systems with transiting planets: WASP-92, WASP-93 and WASP-118. Using ground-based photometric observations of WASP-92 and WASP-93 and \textit{Kepler-K2} observations of WASP-118, we redetermined the orbital and physical parameters of these planets. The precise times of all transits were determined. We constructed O-C diagrams of transits and analysed possible transit timing variations. We did not observe any significant deviation from a linear ephemeris for any of the selected exoplanets. We put upper-mass limits for other hypothetical planets in these systems. Using long-term numerical simulation, we looked for stable regions where another planet could exist for a long time. We used the maximum eccentricity method for this purpose. We discuss the influence of values of initial inclination and eccentricity on the shape and size of regions of stability. 
\end{abstract}


\section{Introduction}

SuperWASP (SWASP) is one of the most successful ground-based photometry projects for searching for extrasolar planets (e.g. \citealt{Kane2008} and \citealt{Hellier2012}). The survey started in 2004 and up to now (26 October 2018), SWASP has discovered 134 exoplanets. It uses two robotic telescopes, one located on the Canary Islands and another one in South Africa. Each one consists of array of eight Canon 200 mm f/1.8 lenses equipped with Andor iKon-L CCD cameras \citep{Pollacco2006}. The large field of view of the lenses gives a big sky coverage of 490 square degrees.

\begin{table}
\caption{The basic parameters of the parent stars of WASP-92 b, WASP-93 b and WASP-118 b (adopted from \citep{Hay2016}, distances from GDR2).}
\label{tab:star-params}
\begin{center}
\begin{tabular}{lccc}
	\hline
	                   &      WASP-92      &      WASP-93      &     WASP-118      \\ \hline
	Spectral type      &        F7         &        F4         &        F6         \\
	Mass [$M_\odot$]   & 1.190 $\pm$ 0.037 & 1.334 $\pm$ 0.033 & 1.320 $\pm$ 0.035 \\
	Radius [$R_\odot$] & 1.341 $\pm$ 0.058 & 1.524 $\pm$ 0.040 & 1.696 $\pm$ 0.029 \\
	V magnitude        &       13.18       &       10.97       &       11.02       \\
	Distance [pc]      &  585.6 $\pm$ 6.6  &  374.0 $\pm$ 7.9  & 380.5 $\pm$ 11.0  \\ \hline
\end{tabular}
\end{center}
\end{table}

For this study we selected three hot Jupiters WASP-92 b, WASP-93 b and WASP-118 b discovered in 2016 and confirmed by \cite{Hay2016}, who used spectral observations to determine the stellar parameters of the parent stars. Distances to these stars are given by the second GAIA Data Release (GDR2) \citep{Gaia2018}. The basic parameters of the parent stars are listed in the Table \ref{tab:star-params}.
Additionally we note that \cite{Mocnik2017} found $\gamma$ Doradus pulsations in the light curve (LC) of WASP-118 with a period 1.9 day and semi-amplitude of $\sim$ 200 ppm.

The aim of this paper was to study the possible presence of other planets in these systems. O-C diagrams of transits were constructed for this purpose and we put upper-mass limits for hypothetical planets in these systems. A second motivation of the present study was to perform simulations leading to the generation of stability maps for the systems.

In Section \ref{obs}, we describe the photometric observations used in this paper. We analyse the LCs we used in Section \ref{lc-analysis}. In Section \ref{ttv}, we search for possible transit timing variations (hereafter TTVs) and study them. Section \ref{stability} is concerned with the long-term stability of these systems. Our results are discussed in Section \ref{disc}.

\section{Observations}
\label{obs}

\begin{figure*}
\includegraphics[width=\linewidth]{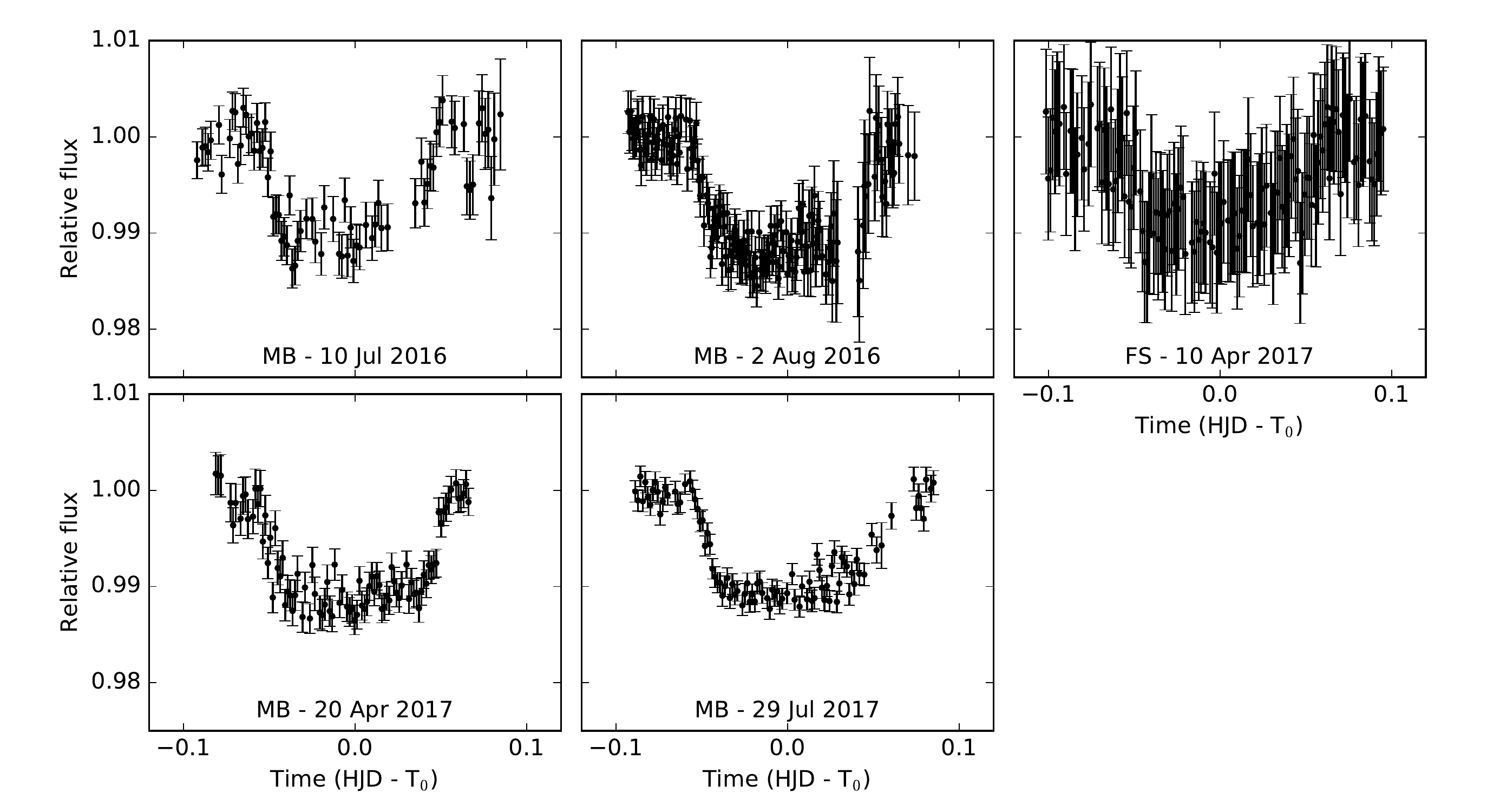}
\caption{Transiting LCs of WASP-92 used for analysis. Detail description in Table \ref{tab:wasp92-obs-time}.}
\label{fig:wasp92-obs}
\end{figure*}

\begin{figure*}
\includegraphics[width=\linewidth]{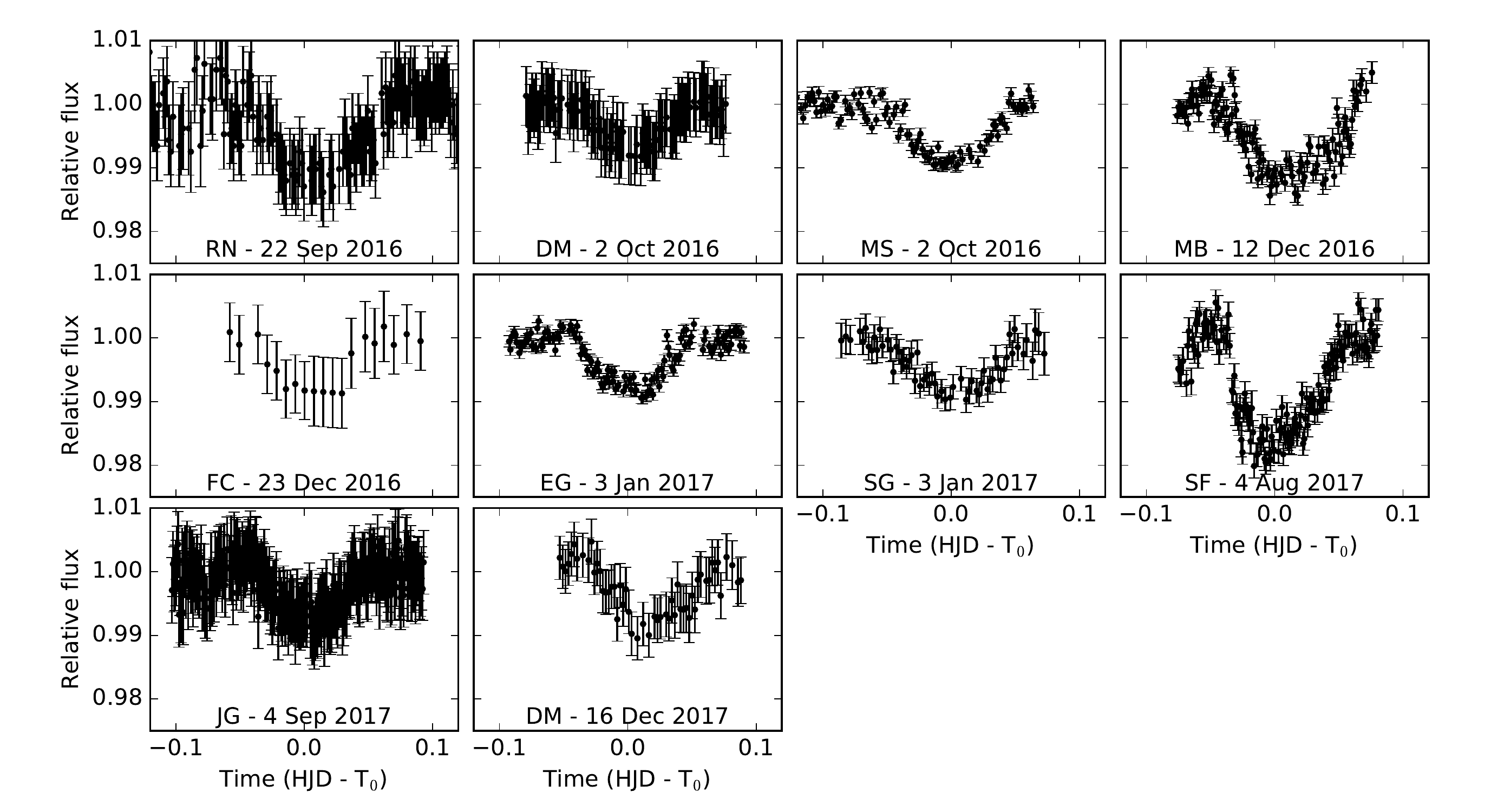}
\caption{Transiting LCs of WASP-93 used for analysis. Detail description in Table \ref{tab:wasp93-obs-time}.}
\label{fig:wasp93-obs}
\end{figure*}

The transiting light curves of WASP-92 and WASP-93 used in this study were taken from the public archive of the Exoplanet Transit Database (ETD)\footnote{http://var2.astro.cz/ETD/} \citep{Poddany2010}. Information about individual observers and the instruments they used is given in Table \ref{tab:obs}. The data from all observations (listed in Tables \ref{tab:wasp92-obs-time} and \ref{tab:wasp93-obs-time} and displayed on Figures \ref{fig:wasp92-obs} and \ref{fig:wasp93-obs}) were used to determine the times of transits of WASP-92 b and WASP-93 b and for TTV analysis.

\begin{table}
\caption{Overview of the telescopes and instruments/detectors used to obtain photometry of WASP-92 and WASP-93. $N_{tr}$ is the number of observed transits. Abbreviations of the observers: 
MB - Marc Bretton (Baronnies Proven\c{c}ales Observatory, France), 
FC - Fran Campos (Observatori Puig d'Agulles, Spain),
SF - St\'{e}phane Ferratfiat (Astronomes  Amateurs  Aixois  de  l'Observatoire  de  Vauvenargues, France),
EG - Eric Girardin (Grand-Pra, Switzerland),
JG - Juanjo Gonzalez (Cielo Profundo, Spain),
SG - Sn\ae{}varr Gu\dh{}mundsson (Nes Observatory, Iceland),
DM - David Molina (Anunaki Observatory, Spain),
RN - Ramon Naves (Observatorio Montcabrer, Spain),
MS - Mark Salisbury (Private Observatory, England),
FS - Francesco Scaggiante (Gruppo Astrofili Salese Galileo Galilei S.M di Sala, Italy).
}
\label{tab:obs}
\begin{center}
\begin{tabular}{cccc}
	\hline
	Observer &     Telescope      &        CCD        & $N_{tr}$ \\ \hline
	   MB    &    Dall-Kirkham    &  SBIG STL-11000   &    2     \\
	         &      430/2940      &                   &  \\
	         & Ritchey Chr\'etien &     FLI PL230     &    3     \\
	         &      820/4780      &                   &  \\	
	   FC    &       Newton       &   SBIG ST-8XME    &    1     \\
	         &      200/940       &                   &  \\
	   SF    &     Cassegrain     &    QSI 583WSG     &    1     \\
	         &      350/3910      &                   &  \\	         
	   EG    & Ritchey Chr\'etien &  SBIG STL-11000   &    1     \\
	         &      400/3360      &                   &  \\
	   JG    &     Cassegrain     &   SBIG ST8-XME    &    1     \\
	         &      235/1645      &                   &  \\	         
	   SG    &     Cassegrain     &  SBIG STL-11000   &    1     \\
	         &      300/3000      &                   &  \\   
	   DM    &     Cassegrain     &   SBIG ST8-XME    &    2     \\
	         &      200/2000      &                   &  \\
	   RN    &     Cassegrain     &    MI G4-9000     &    1     \\
	         &      300/3000      &                   &  \\
	   MS    &    Dall-Kirkham    &   SBIG ST10-XME   &    1     \\
	         &      400/2720      &                   &  \\
	   FS    &       Newton       & Audine KAF1600 ME &    1     \\
	         &      410/1710      &                   &  \\
 \hline
\end{tabular}
\end{center}
\end{table}

\begin{table}
\caption{Summary of the observing runs for WASP-92. Observers according to Table \ref{tab:obs}. The dates are given for the center of transit. $N_{exp}$ – number of useful exposures, $t_{exp}$ – exposure time.}
\label{tab:wasp92-obs-time}
\begin{center}
\begin{tabular}{ccccc}
	\hline
	Observer &    Date     & Filter & $N_{exp}$ & $t_{exp}$ (s) \\ \hline
	MB (DK)  & 20 Jul 2016 & Clear  &    86     &      120      \\
	MB (RC)  & 2 Aug 2016  & Clear  &    196    &      60       \\
	   FS    & 10 Aug 2017 &   R    &    145    &      90       \\
	MB (RC)  & 20 Apr 2017 &   I    &    101    &      120      \\
	MB (RC)  & 29 Jul 2017 &   I    &    105    &      120      \\ \hline
\end{tabular}
\end{center}
\end{table}

\begin{table}
\caption{Summary of the observing runs for WASP-93. For a detailed description see Table \ref{tab:wasp92-obs-time}.}
\label{tab:wasp93-obs-time}
\begin{center}
\begin{tabular}{ccccc}
	\hline
	Observer &    Date     & Filter & $N_{exp}$ & $t_{exp}$ (s) \\ \hline
	   RN    & 22 Sep 2016 &   r'   &    154    &      100      \\
	   DM    & 2 Oct 2016  & Clear  &    102    &      120      \\
	   MS    & 2 Oct 2016  &   R    &    110    &      30       \\
	MB (DK)  & 12 Dec 2016 & Clear  &    131    &      60       \\
	   FC    & 23 Dec 2016 &   R    &    20     &      600      \\
	   EG    & 3 Jan 2017  & Clear  &    137    &      90       \\
	   SG    & 3 Jan 2017  & Clear  &    71     &      120      \\
	   SF    & 4 Aug 2017  &   R    &    158    &      60       \\
	   JG    & 4 Sep 2017  & Clear  &    252    &      60       \\
	   DM    & 16 Dec 2017 & Clear  &    63     &      180      \\ \hline
\end{tabular}
\end{center}
\end{table}

There were no observations of WASP-118 available in the ETD. However, this object was observed during Campaign 8 of the \textit{Kepler K2} mission \citep{Howell2014}. Those observations started on 3 January 2016 and finished on 23 March 2016. We used  de-trended short-cadence data (PDCSAP\_FLUX) which are sampled every 58.8 seconds. The data are available to the public at the Mikulski Archive for Space Telescopes (MAST)\footnote{https://archive.stsci.edu/index.html, doi:10.17909/T9TG7J}. During the observation period 19 transits of WASP-118 b were observed.

\section{Light curve analysis}
\label{lc-analysis}

As a first step of our LC analysis, we fitted the out-of-transit parts of the LCs by a second-order polynomial function to remove additional long-term trends and to normalize the LC levels. 

In all cases, we used our software implementation of \citeauthor{Mandel2002}'s \citeyearpar{Mandel2002} model for fitting transits. Markov chain Monte Carlo (MCMC) simulation was used to determine transit parameters. This method takes into account individual errors of observations and gives a realistic and statistically significant estimate of parameter values and errors. As a starting point for the MCMC fitting, we used the parameters of the planet given in \citet{Hay2016}. We have used a quadratic model of limb darkening with coefficients from \cite{Claret2013} selected according to filter. We ran the MCMC simulation with 10$^6$ steps.

\subsection{WASP-92 and WASP-93}

For the ground-based observations of WASP-92 and WASP-93, we selected those observations with the best signal-to-noise ratio as templates. These templates were fitted using the MCMC method as mentioned above. 

\begin{figure}
\includegraphics[width=\linewidth]{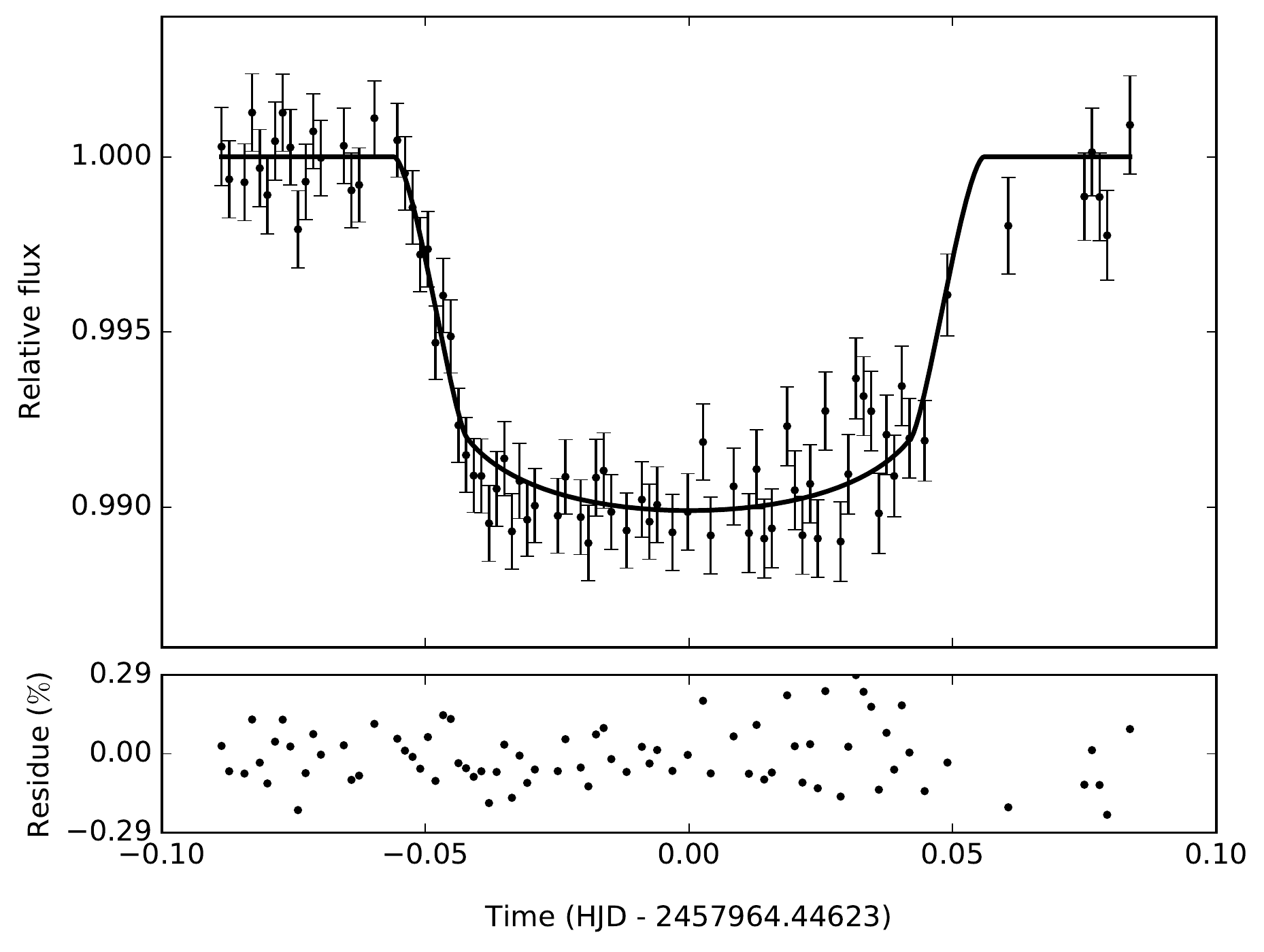}
\includegraphics[width=\linewidth]{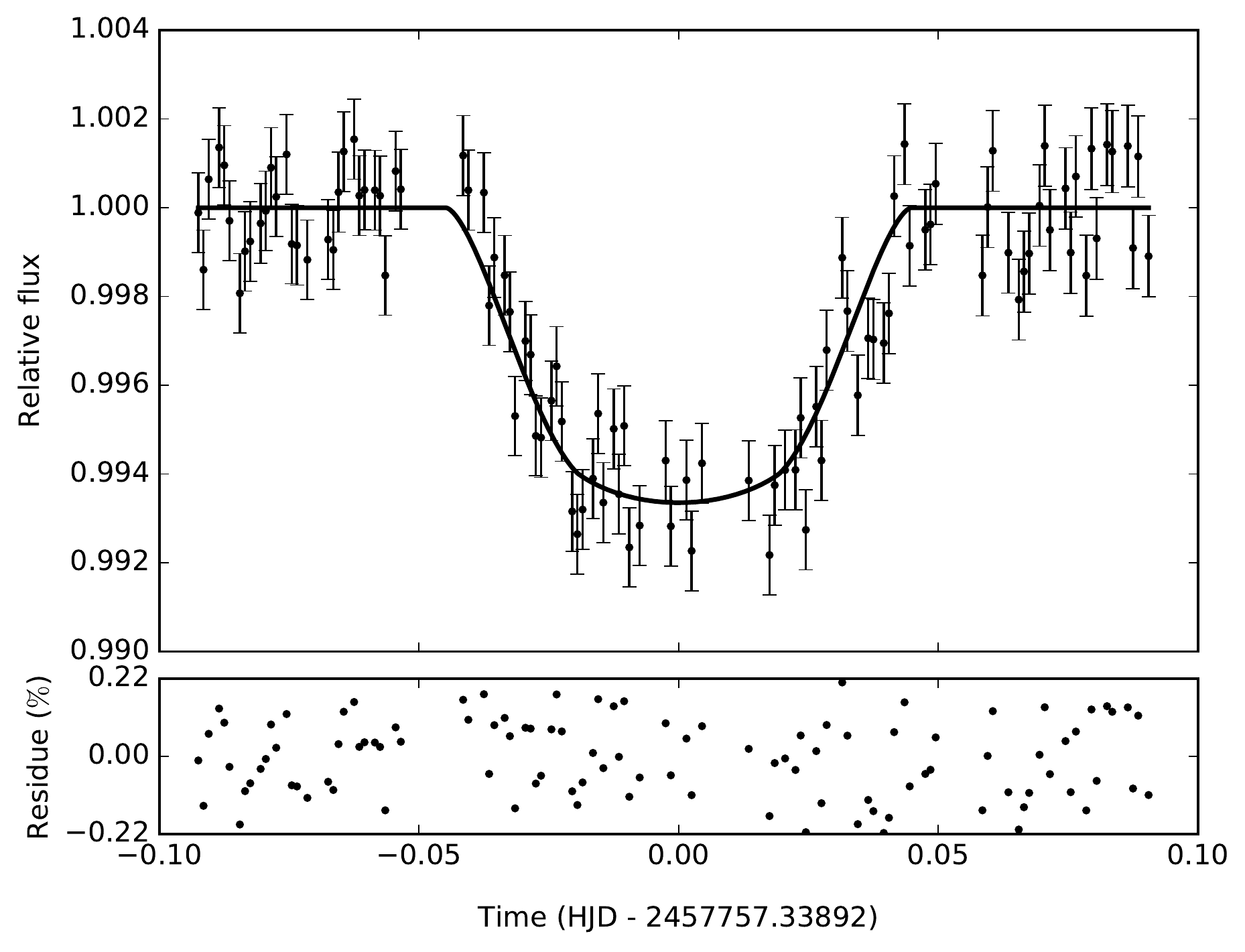}
\includegraphics[width=\linewidth]{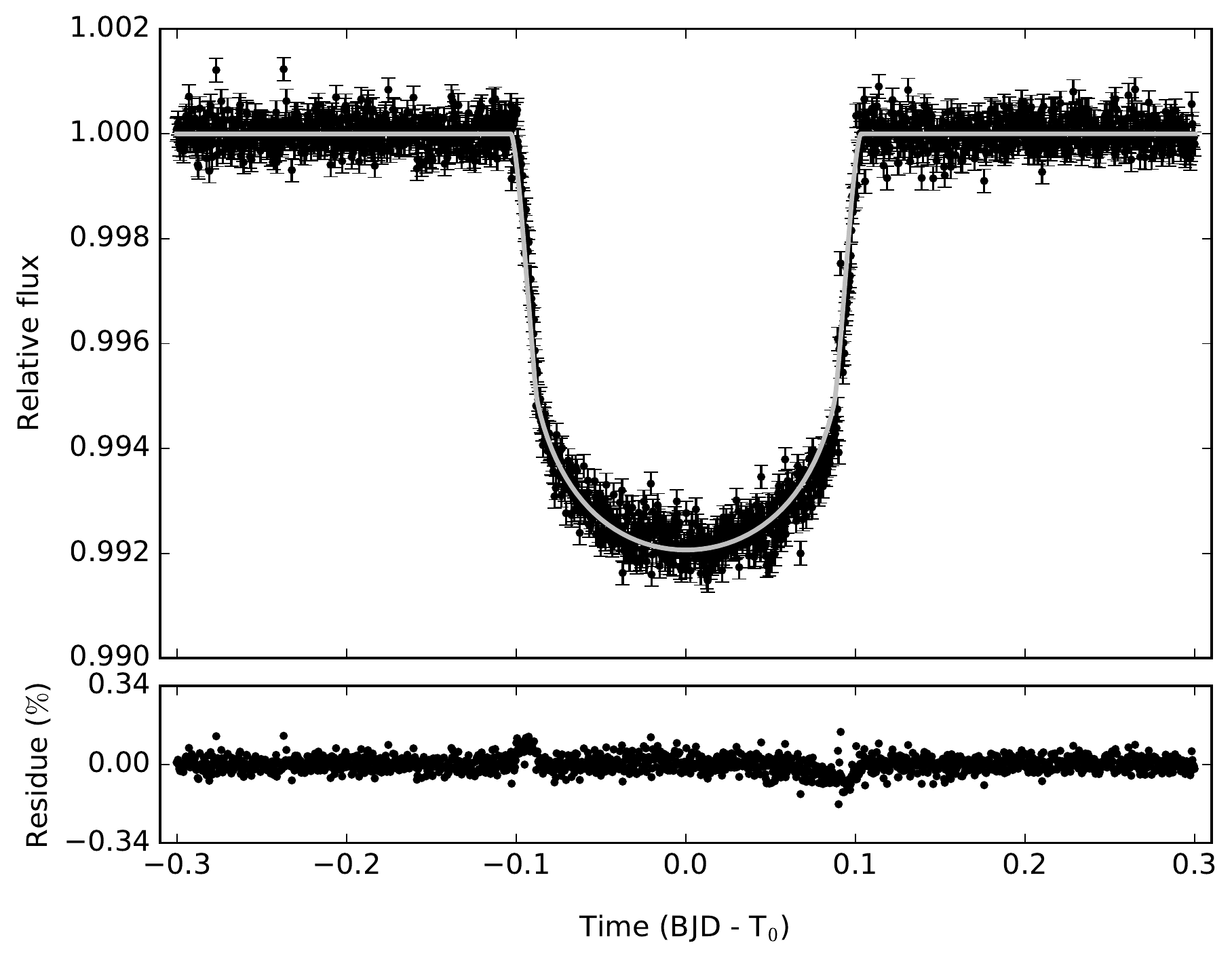}
\caption[]{\textit{Top: }Template of the transit of WASP-92 b (based on Marc Bretton's observations from 29 July 2017). The solid line marks the best fit of the transit model with the parameters in Table \ref{tab:wasp92-params}.\\ \textit{Middle: } Template of the transit of WASP-93 b (based on Eric Girardin's observations from 3 January 2017). The solid line marks the best fit of the transit model with the parameters in Table \ref{tab:wasp93-params}.\\ \textit{Bottom: }Template of the transit of WASP-118 b. Only every tenth data point is displayed on the figure. The solid line marks the best fit of the transit model with the parameters in Table \ref{tab:wasp118-params}.}
\label{fig:template}
\end{figure}

\begin{table}
\caption{Parameters of exoplanet WASP-92 b, $T_0$ - transit epoch (in HJD - 2450000), $P$ - orbital period, $a$ - semi-major axis of the planetary orbit, $r_{\rm p}$ - planet radius, $i$ - orbital inclination, $\Delta$ - depth of transit, $t_{\rm T}$ - transit duration, $b$ - impact parameter, $\chi^2$ - sum of squares of the best fit, $\chi^2/n$ - reduced sum of squares. The linear ephemeris ($T_0$ and $P$) was obtained from an analysis of the times of transit (see Section \ref{ttv}).}
\label{tab:wasp92-params}
\begin{center}
\begin{tabular}{lcc}
	\hline
	Parameter                  &     \cite{Hay2016}      &       This paper        \\ \hline
	$T_0$ [HJD]                & 6381.28340$\pm$0.00027  & 6381.26836$\pm$0.00310  \\
	$P$ [d]                    & 2.1746742$\pm$0.0000016 & 2.1746949$\pm$0.0000046 \\
	$a$ [au]                   &   0.03480$\pm$0.00036   &   0.03684$\pm$0.00358   \\
	$a$ [$R_\star$]            &      5.58$\pm$0.26      &      5.91$\pm$0.51      \\
	$r_{\rm p}$ [$R_{\oplus}$] &    16.376$\pm$0.863     &    14.071$\pm$0.676     \\
	$r_{\rm p}$ [$R_\star$]    &    0.1121$\pm$0.0077    &    0.0963$\pm$0.0017    \\
	$i$ [\degr]                &     83.75$\pm$0.69      &     84.65$\pm$1.81      \\ \hline
	$\Delta$                   &   0.01254$\pm$0.00172   &   0.00927$\pm$0.00033   \\
	$t_{\rm T}$ [d]            &    0.1153$\pm$0.0012    &    0.1113$\pm$0.0183    \\
	$b$ [$R_\star$]            &     0.608$\pm$0.043     &     0.552$\pm$0.192     \\ \hline
	$\chi^2$                   &         473.81          &          73.95          \\
	$\chi^2/n$                 &          5.92           &          0.92           \\ \hline
\end{tabular}
\end{center}
\end{table}

To determine the new parameters of WASP-92 b, we have selected Marc Bretton's observations from 29 July 2017 (on Figure \ref{fig:template}  (\textit{top})). For WASP-92 b, the results of our LC analysis show that the radius of this exoplanet is significantly smaller than the value given by \cite{Hay2016}. All the parameters of planet are shown in Table \ref{tab:wasp92-params}. The slightly larger uncertainties of some parameters could be caused by the poorer quality and greater variance of ground-based observations.

\begin{table}
\caption{Parameters of exoplanet WASP-93 b. For a detailed description see Table \ref{tab:wasp92-params}.}
\label{tab:wasp93-params}
\begin{center}
\begin{tabular}{lcc}
	\hline
	Parameter                  &     \cite{Hay2016}      &       This paper        \\ \hline
	$T_0$ [HJD]                & 6079.56420$\pm$0.00045  & 6079.55280$\pm$0.00457  \\
	$P$ [d]                    & 2.7325321$\pm$0.0000020 & 2.7325559$\pm$0.0000073 \\
	$a$ [au]                   &   0.04211$\pm$0.00035   &   0.04559$\pm$0.00275   \\
	$a$ [$R_\star$]            &      5.96$\pm$0.16      &      6.45$\pm$0.35      \\
	$r_{\rm p}$ [$R_{\oplus}$] &    17.901$\pm$0.863     &    14.472$\pm$0.568     \\
	$r_{\rm p}$ [$R_\star$]    &    0.1080$\pm$0.0059    &    0.0873$\pm$0.0025    \\
	$i$ [\degr]                &     81.18$\pm$0.29      &     82.27$\pm$0.58      \\ \hline
	$\Delta$                   &   0.01097$\pm$0.00013   &   0.00762$\pm$0.00044   \\
	$t_{\rm T}$ [d]            &    0.0931$\pm$0.0010    &    0.0885$\pm$0.0180    \\
	$b$ [$R_\star$]            &     0.904$\pm$0.009     &     0.868$\pm$0.080     \\ \hline
	$\chi^2$                   &         213.98          &         150.74          \\
	$\chi^2/n$                 &          2.04           &          1.44           \\ \hline
\end{tabular}
\end{center}
\end{table}

Eric Girardin's observations from 3 January 2017 (on Figure \ref{fig:template} (\textit{middle})) were selected in order to determine the parameters of WASP-93 b. The parameters of this planet are listed in Table \ref{tab:wasp93-params}. Our results indicate a smaller exoplanet than found by \cite{Hay2016} results.

\subsection{WASP-118}

Because of the high quality of \textit{Kepler-K2} data, a different approach was used for WASP-118 b.
For this planet we aligned and stacked all 19 LCs together and this stacked LC was fitted by MCMC method as described above. This is similar to the method we used in our TTV analysis of Kepler-410~A~b in our paper \cite{Gajdos2017}.  

The template of the WASP-118 b transit together with the model LC is displayed on Figure \ref{fig:template}  (\textit{bottom}). We compared our parameters with those from papers by \cite{Hay2016} and \cite{Mocnik2017} in a Table \ref{tab:wasp118-params}. As expected, our results are closer to the values given by \cite{Mocnik2017} because they also used Kepler-K2 data and \cite{Hay2016} worked with different data.

\begin{table*}
\caption{Parameters of exoplanet WASP-118 b. For a detailed description see Table \ref{tab:wasp92-params}.}
\label{tab:wasp118-params}
\begin{center}
\begin{tabular}{lccc}
	\hline
	Parameter                  &     \cite{Hay2016}      &    \cite{Mocnik2017}     &       This paper        \\ \hline
	$T_0$ [BJD]                & 6787.81423$\pm$0.00062  & 6787.81256$^*\pm$0.00002 & 6787.81249$\pm$0.00068  \\
	$P$ [d]                    & 4.0460435$\pm$0.0000044 & 4.0460407$\pm$0.0000026  & 4.0460654$\pm$0.0000043 \\
	$a$ [au]                   &   0.05453$\pm$0.00048   &   0.05450$\pm$0.00049    &   0.05356$\pm$0.00099   \\
	$a$ [$R_\star$]            &     6.899$\pm$0.136     &     6.678$\pm$0.060      &     6.776$\pm$0.006     \\
	$r_{\rm p}$ [$R_{\oplus}$] &    16.141$\pm$0.404     &     15.630$\pm$0.150     &    14.940$\pm$0.268     \\
	$r_{\rm p}$ [$R_\star$]    &   0.08707$\pm$0.00266   &   0.08166$\pm$0.00153    &   0.08059$\pm$0.00003   \\
	$i$ [\degr]                &     88.70$\pm$0.90      &      88.24$\pm$0.14      &     89.86$\pm$0.18      \\ \hline
	$\Delta$                   &   0.00755$\pm$0.00019   &   0.00668$\pm$0.00001    &   0.00649$\pm$0.00001   \\
	$t_{\rm T}$ [d]            &    0.2002$\pm$0.0019    &    0.2046$\pm$0.0001     &    0.2062$\pm$0.0002    \\
	$b$ [$R_\star$]            &     0.160$\pm$0.100     &     0.206$\pm$0.015      &     0.017$\pm$0.021     \\ \hline
	$\chi^2$                   &        116249.01        &         19040.62         &        17746.10         \\
	$\chi^2/n$                 &          8.55           &           1.40           &          1.31           \\ \hline
\end{tabular}
\end{center}
\begin{flushleft}
$^*$ original value 7423.04483 shifted to the same epoch
\end{flushleft}
\end{table*}

\section{Transit timing variations}
\label{ttv}

We have repeated the MCMC simulation using the solution of our template LC (see Section \ref{lc-analysis}) as the starting point for each individual transit interval, letting only the time of transit $T_{\rm T}$ update. Values of $T_{\rm T}$ were used to construct an O-C diagram.

\begin{figure}
\includegraphics[width=\linewidth]{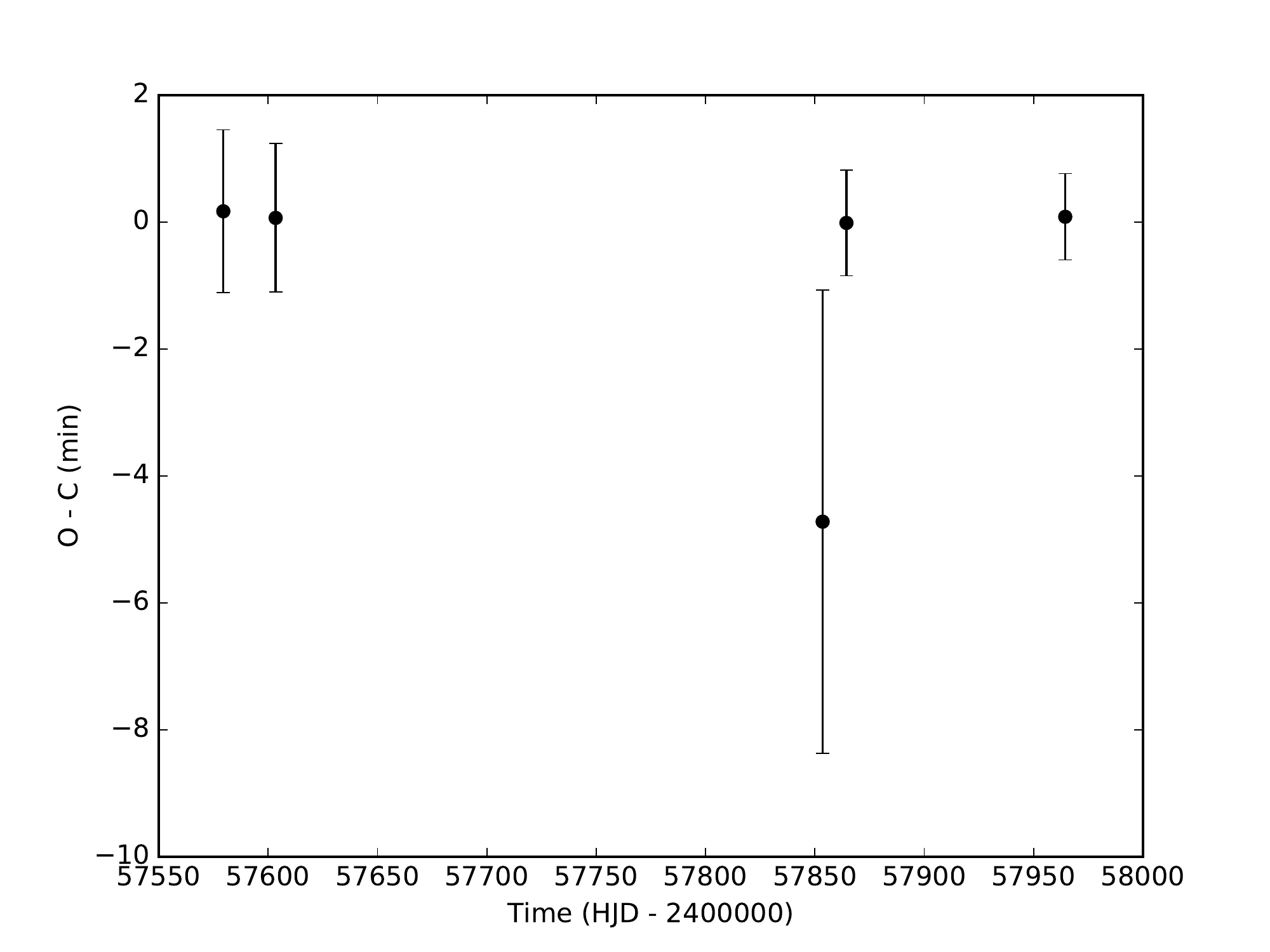}
\includegraphics[width=\linewidth]{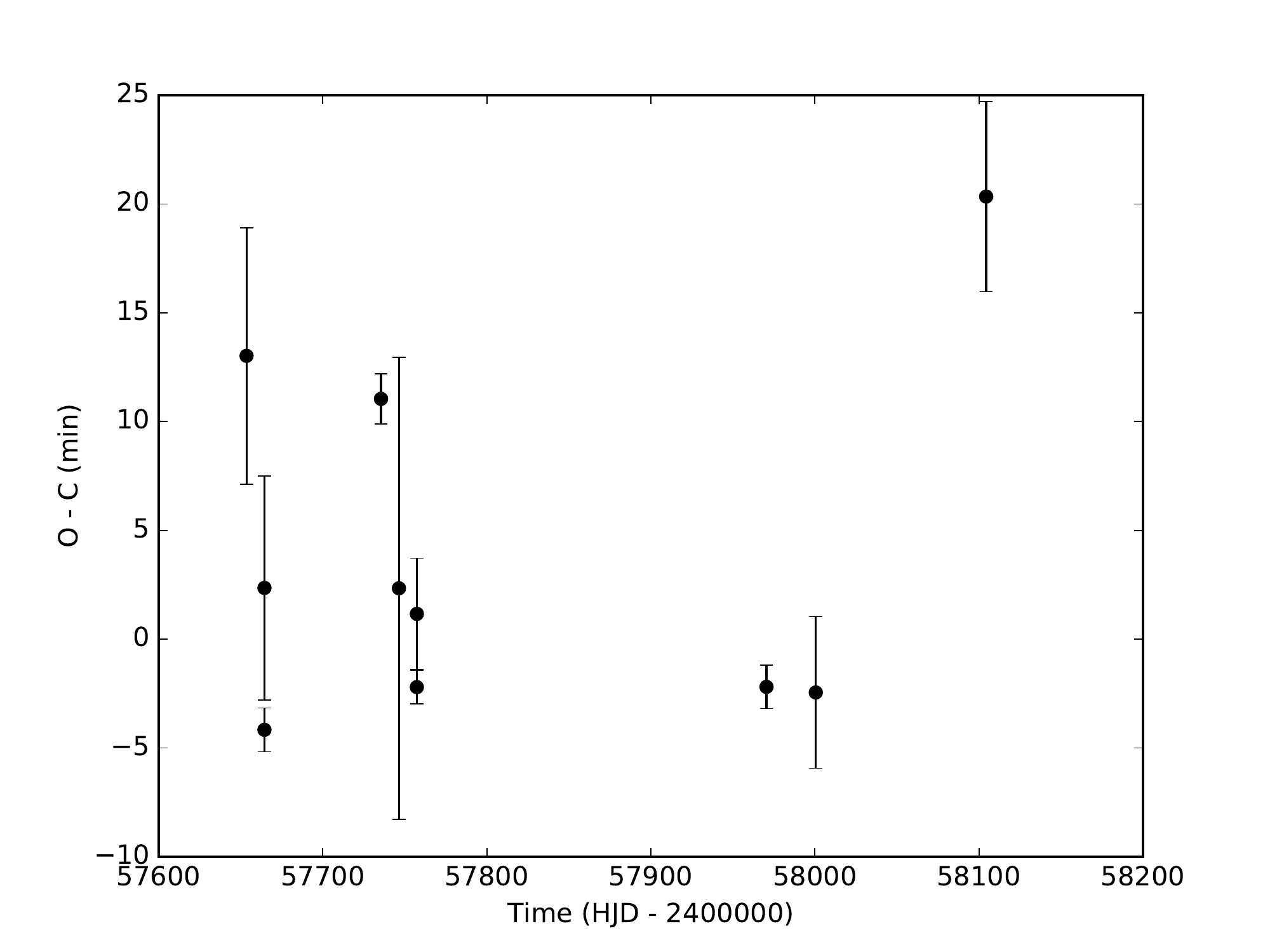}
\includegraphics[width=\linewidth]{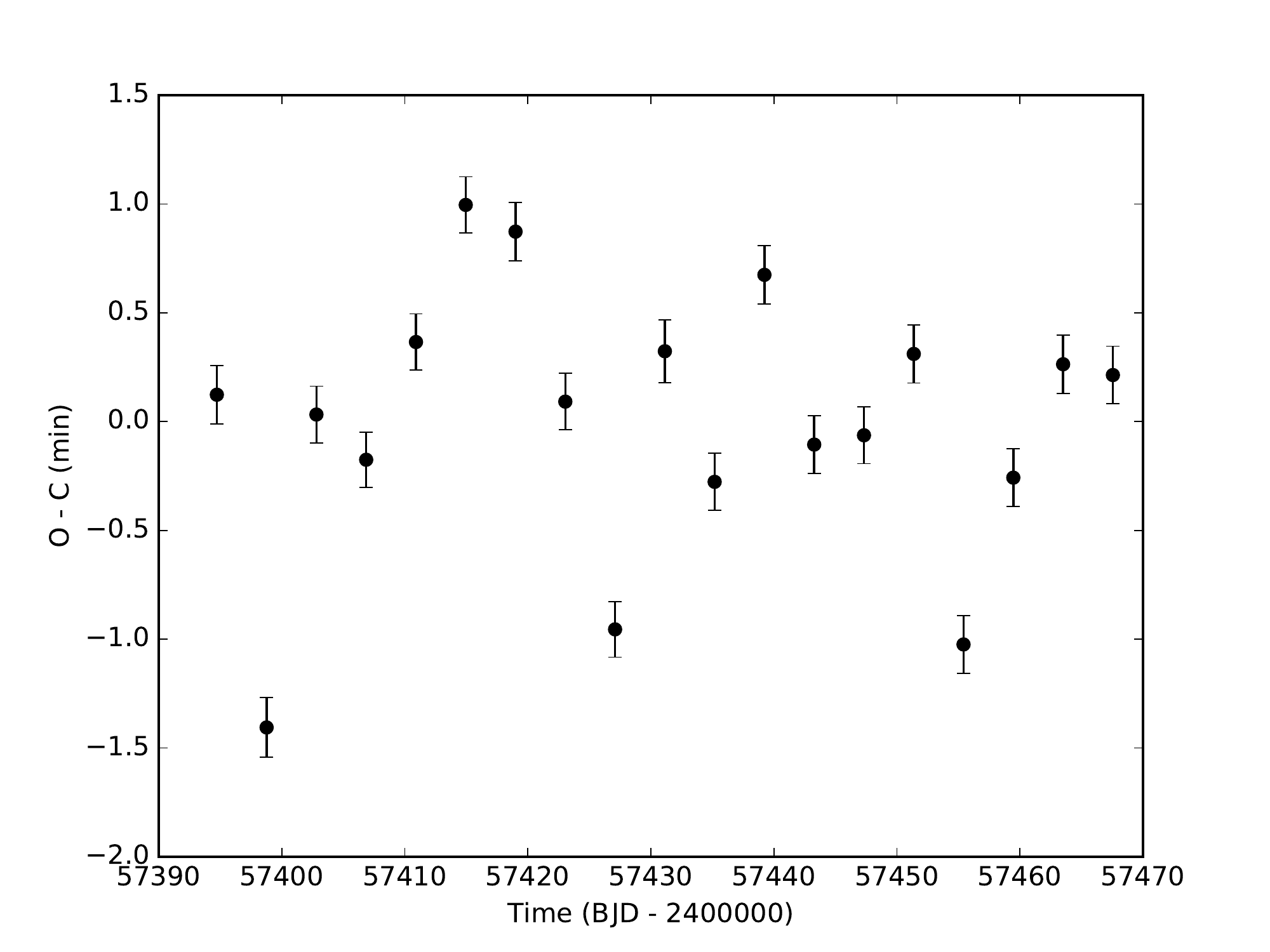}
\caption{O-C diagrams for transit timing of WASP-92 b (\textit{top}), WASP-93 b (\textit{middle}) and WASP-118 b (\textit{bottom}), plotted according to the new linear ephemeris.}
\label{fig:ttv}
\end{figure}

No significant periodic variations were observed on any O-C transit diagram. The times of transits obtained were fitted with a linear function using the MCMC method. The calculated values of period and transit epoch, with their uncertainties, are listed in Tables \ref{tab:wasp92-params}, \ref{tab:wasp93-params} and \ref{tab:wasp118-params}. O-C diagrams of transits calculated according to the new linear ephemerides are in Figure \ref{fig:ttv}. O-C diagrams of ground-based observations of WASP-92 and WASP-93 are much noisier than the \textit{Kepler-K2} observation of WASP-118. For these noisy cases, any periodic TTV signal with amplitude greater than the maximum dispersion of these O-C diagrams over the observing period is unlikely to be seen.

Following \cite{Gibson2009}, we put upper constraints on the mass of a potential perturbing planet in the system with our refined assumptions. We set the maximum variance on our O-C diagram as the amplitude of possible TTVs. We assumed that the orbits of both planets are circular and coplanar. The parameters of the transiting planet were taken from \cite{Hay2016}. The period of a perturbing planet varied from 0.2 to 5 times the orbital period of the transiting planet. 

We used Chamber's \textsf{MERCURY6} code \citep{Chambers1999} to produce 20000 synthetic O-C diagrams for different configurations of mass and orbital period of a hypothetical planet. We applied the Bulirsch-Stoer algorithm to solve our three-body problem. Our calculations covered the whole length of the observing period which, for the ground-based observations of WASP-92 and WASP-93 did not exceed two years, and was only three months for the \textit{Kepler-K2} observations of WASP-118.

\begin{figure}
\includegraphics[width=\linewidth]{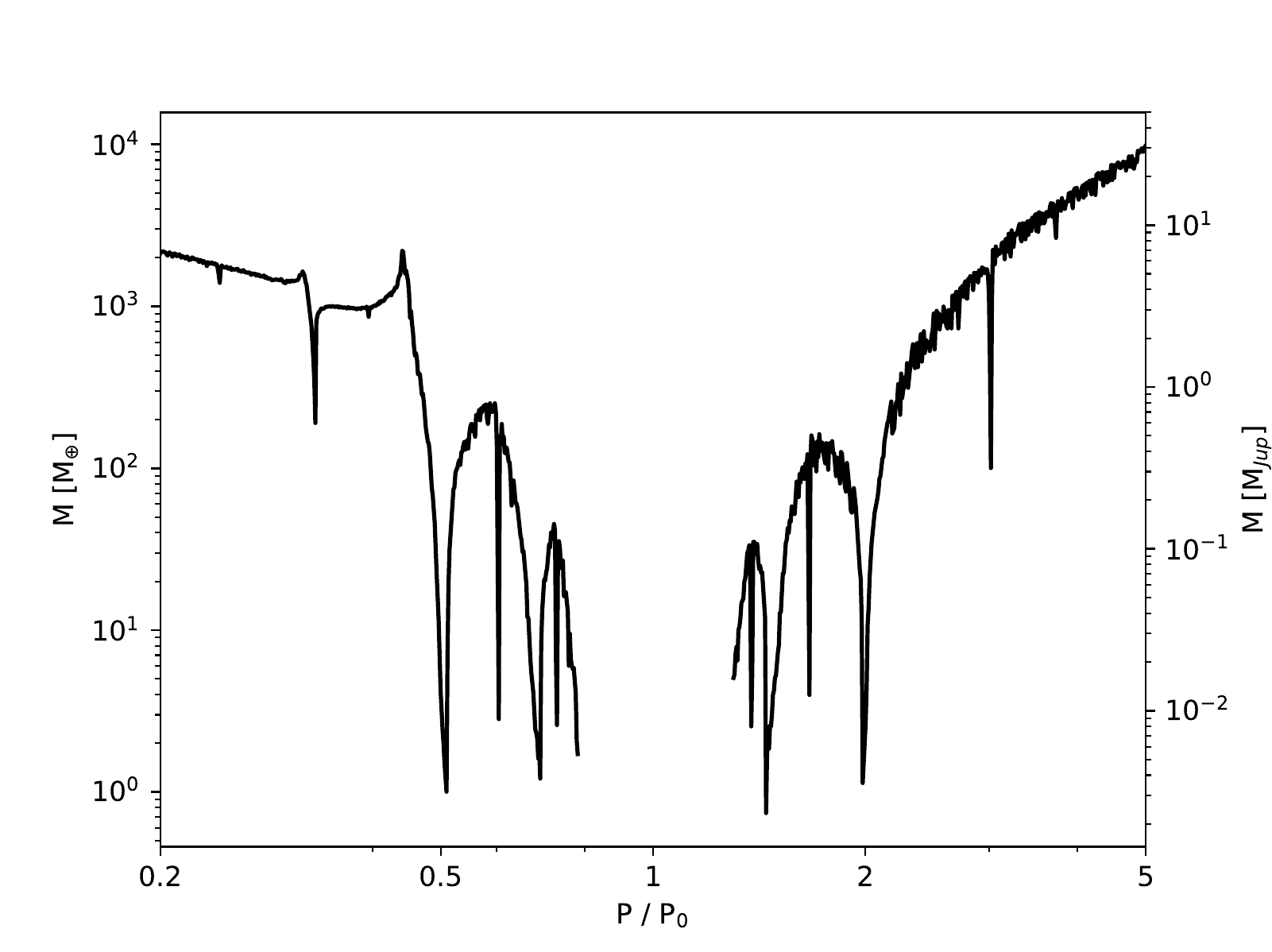}
\includegraphics[width=\linewidth]{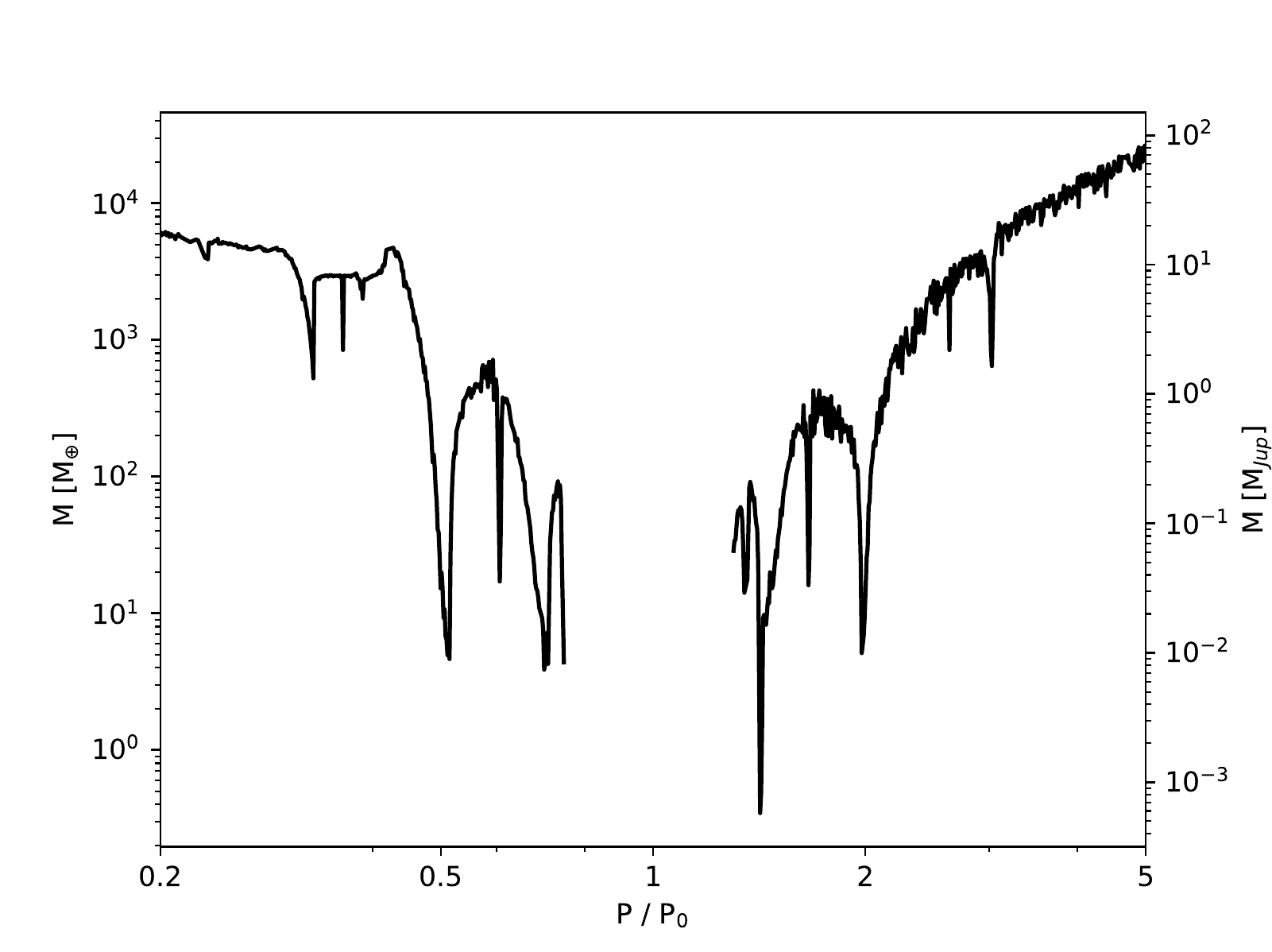}
\includegraphics[width=\linewidth]{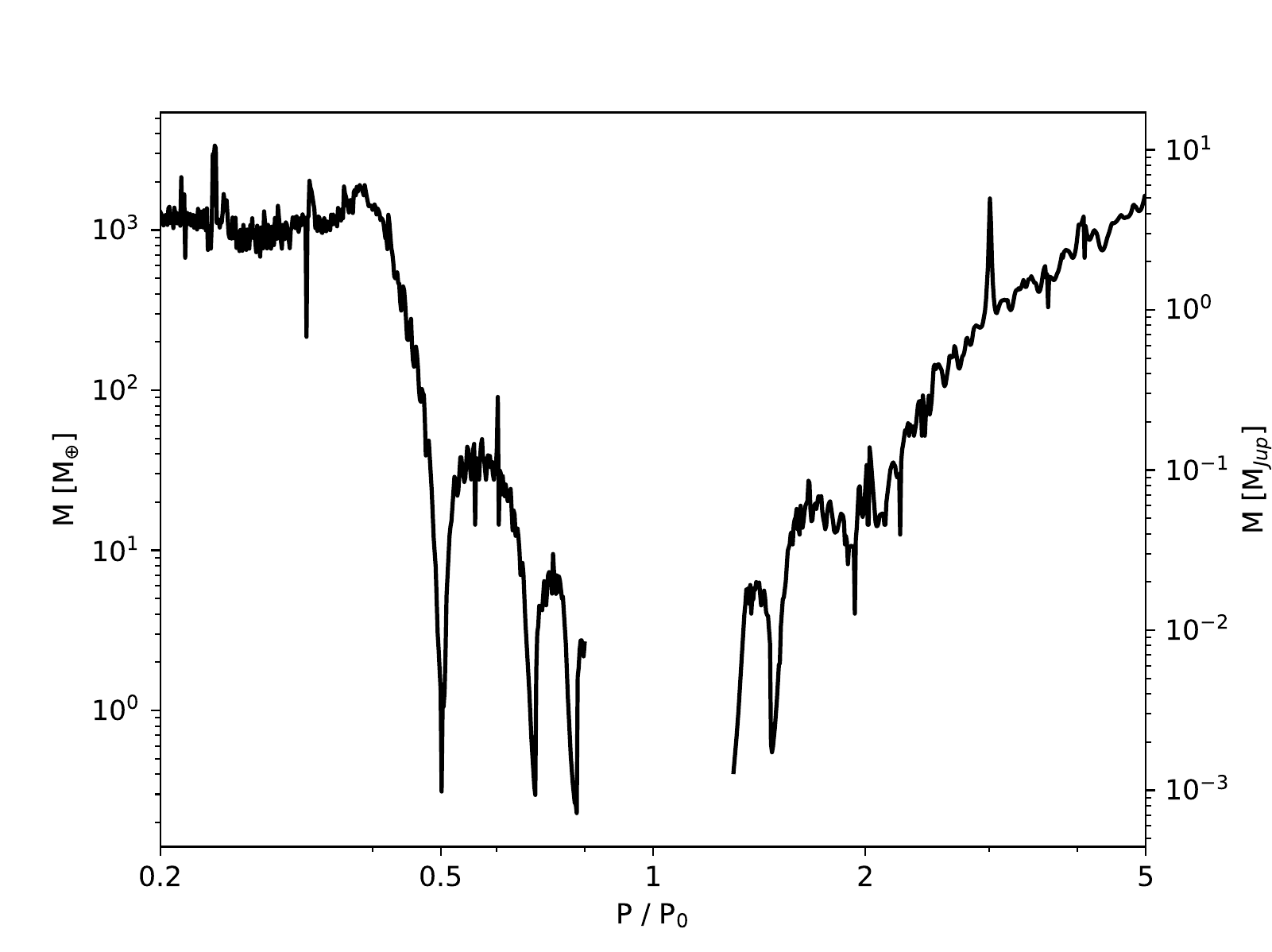}
\caption{The upper-mass limits of a hypothetical additional planet in the WASP-92 system (\textit{top}), WASP-93 system (\textit{middle}) and WASP-118 system (\textit{bottom}).}
\label{fig:limit}
\end{figure}

The results of our simulations are on Figure \ref{fig:limit}. They display the upper-mass limit of a hypothetical perturbing planet as a function of the ratio of the orbital period of the transiting planet $P_0$ and that of the perturbing planet $P$. Regions around $P/P_0 \approx 1 $ were found to be unstable due to frequent close encounters of the planets. The general funnel-like shape of these limits is given by the condition of Hill stability \citep{Barnes2006}. For orbits with a period ratio $P/P_0 \lesssim 0.4$, the tangential motion of the parent star caused by additional planet became the dominant source of TTV and thus the mass limit is linear (nearly constant) in logarithmic scale in this region. Its behaviour is similar to the upper-mass limit calculated from astrometric observations \citep{Agol2005}. Some mean-motion resonances (MMR) are clearly visible. The strongest MMR are inner resonances 1:3 (not visible for WASP-118), 1:2, 2:3 and 3:4 (only in the WASP-118 system) and also exterior MMR 3:2, 2:1 (weak for WASP-118) and 3:1 (not visible in the WASP-118 system; very narrow for WASP-92). The small increase of upper-mass limit in the region $P/P_0 \approx 0.4 - 0.45$ was also observed in all studied systems. A similar feature in this region was found but not discussed in few other papers \citep[e.g.][]{Gibson2009}. Our analysis also allows the existence of an Earth-mass or super-Earth-mass planet close to the MMR. A longer observing span and better precision of the observations would allow us to specify lower upper-mass limits for our hypothetical planets.

\section{Long-term stability of the systems}
\label{stability}

\begin{figure}
\includegraphics[width=\linewidth]{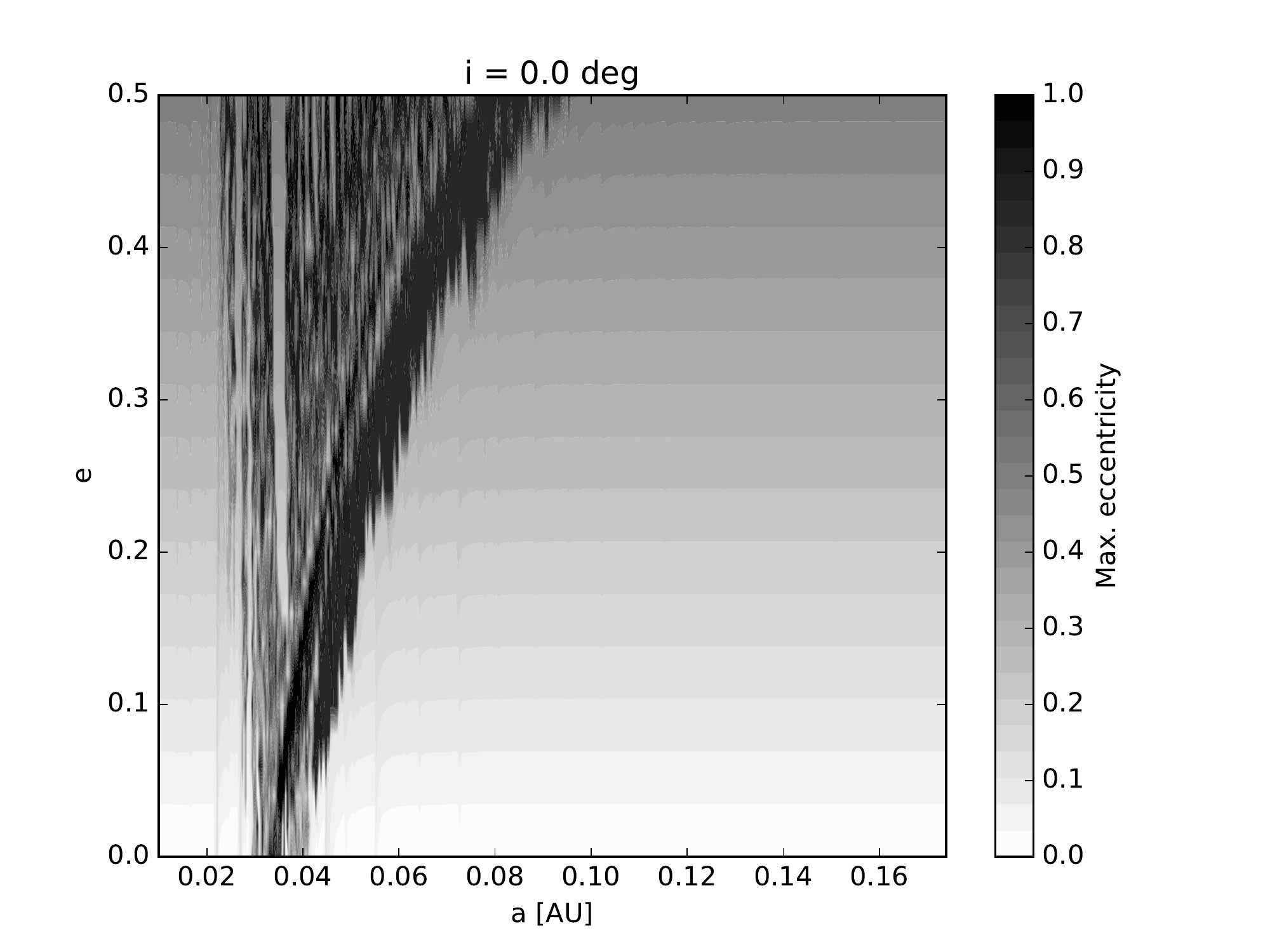}
\includegraphics[width=\linewidth]{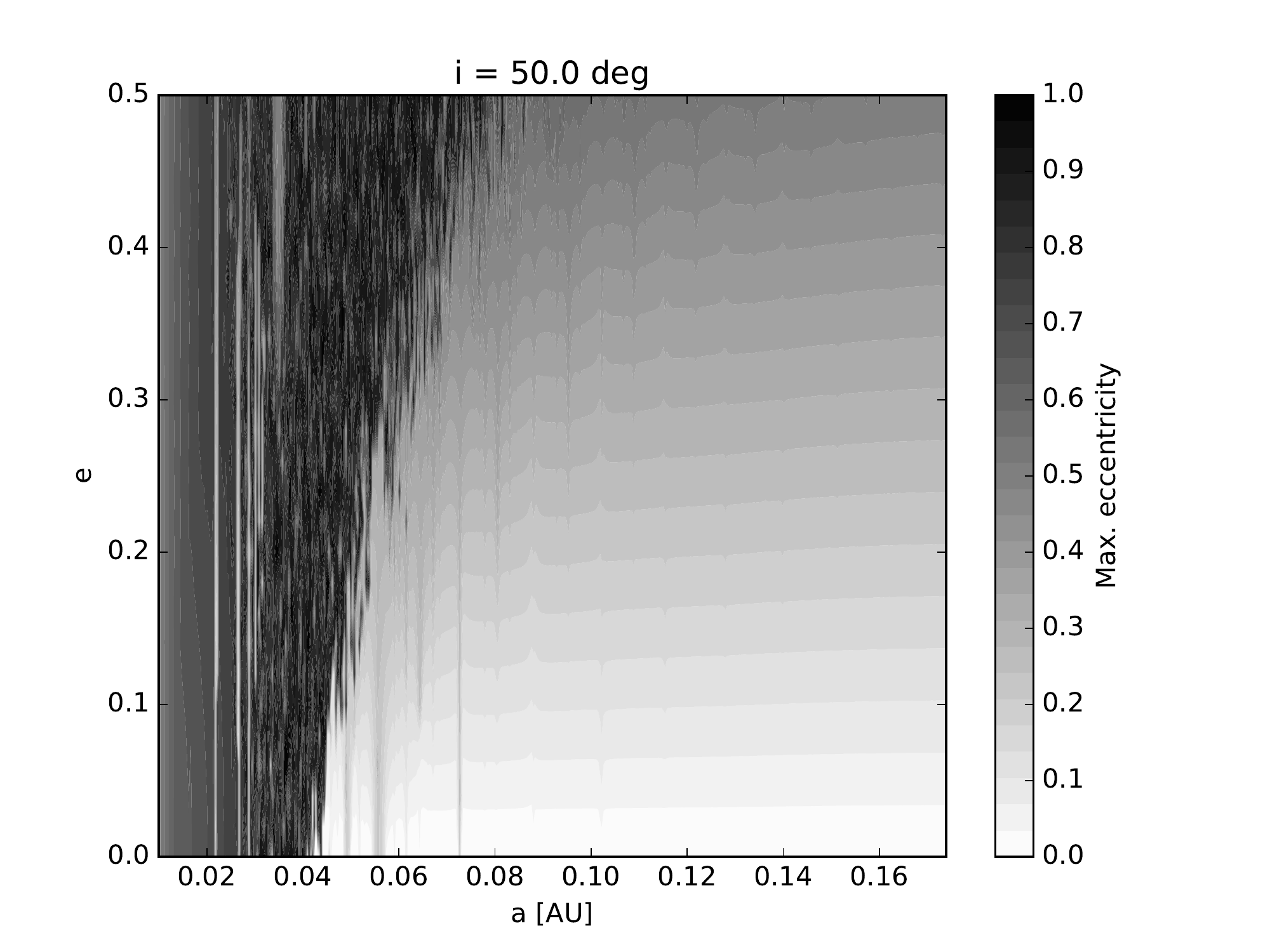}
\includegraphics[width=\linewidth]{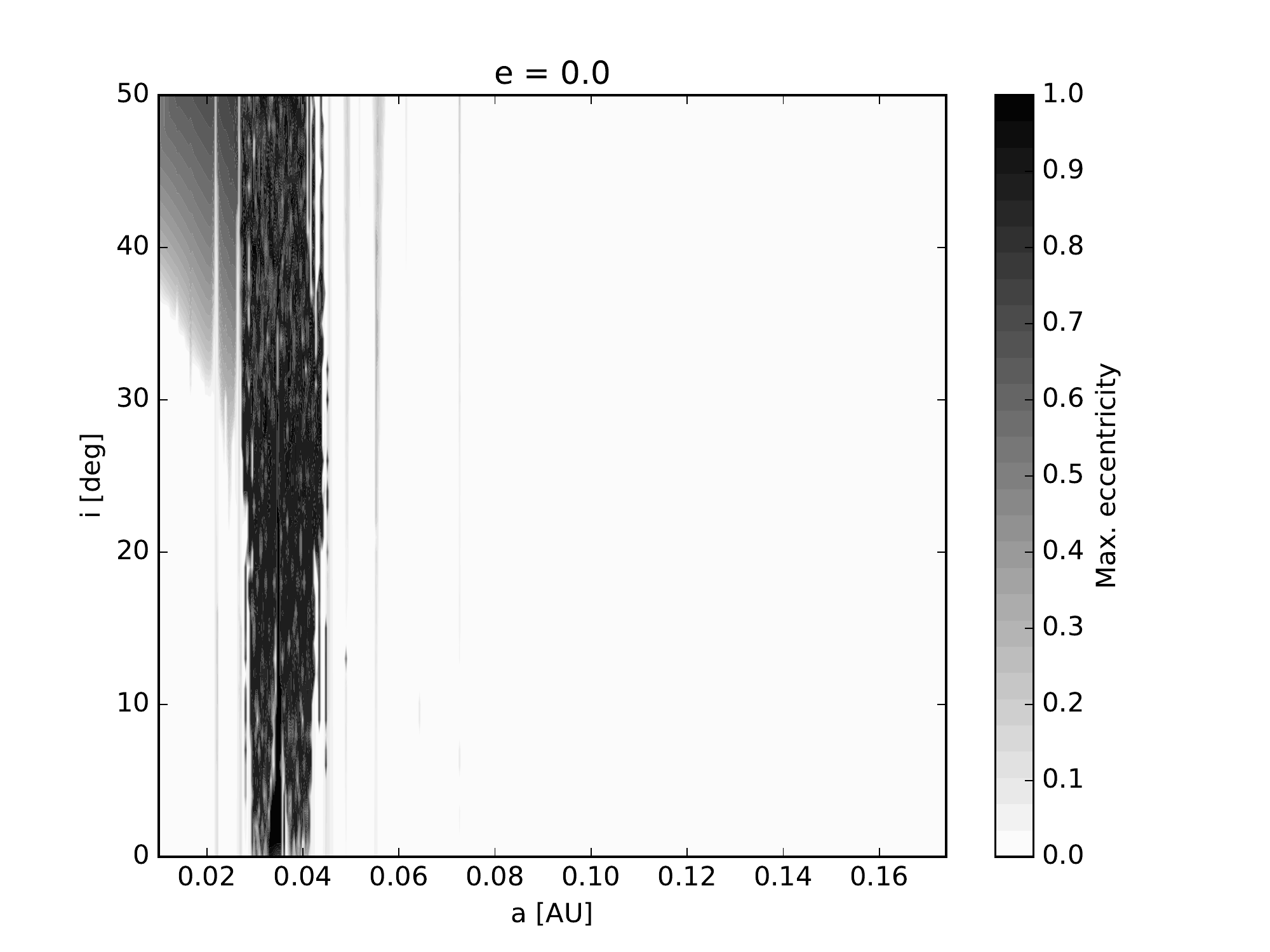}
\caption{Stability plot showing the maximum eccentricity for the WASP-92 system in the $a - e$ plane for $i = 0\degr$ (\textit{top}), for $i = 50\degr$ (\textit{middle}) and in the $a - i$ plane for $e = 0.0$ (\textit{bottom}).}
\label{fig:wasp92-stab}
\end{figure}

\begin{figure*}
\includegraphics[width=\linewidth]{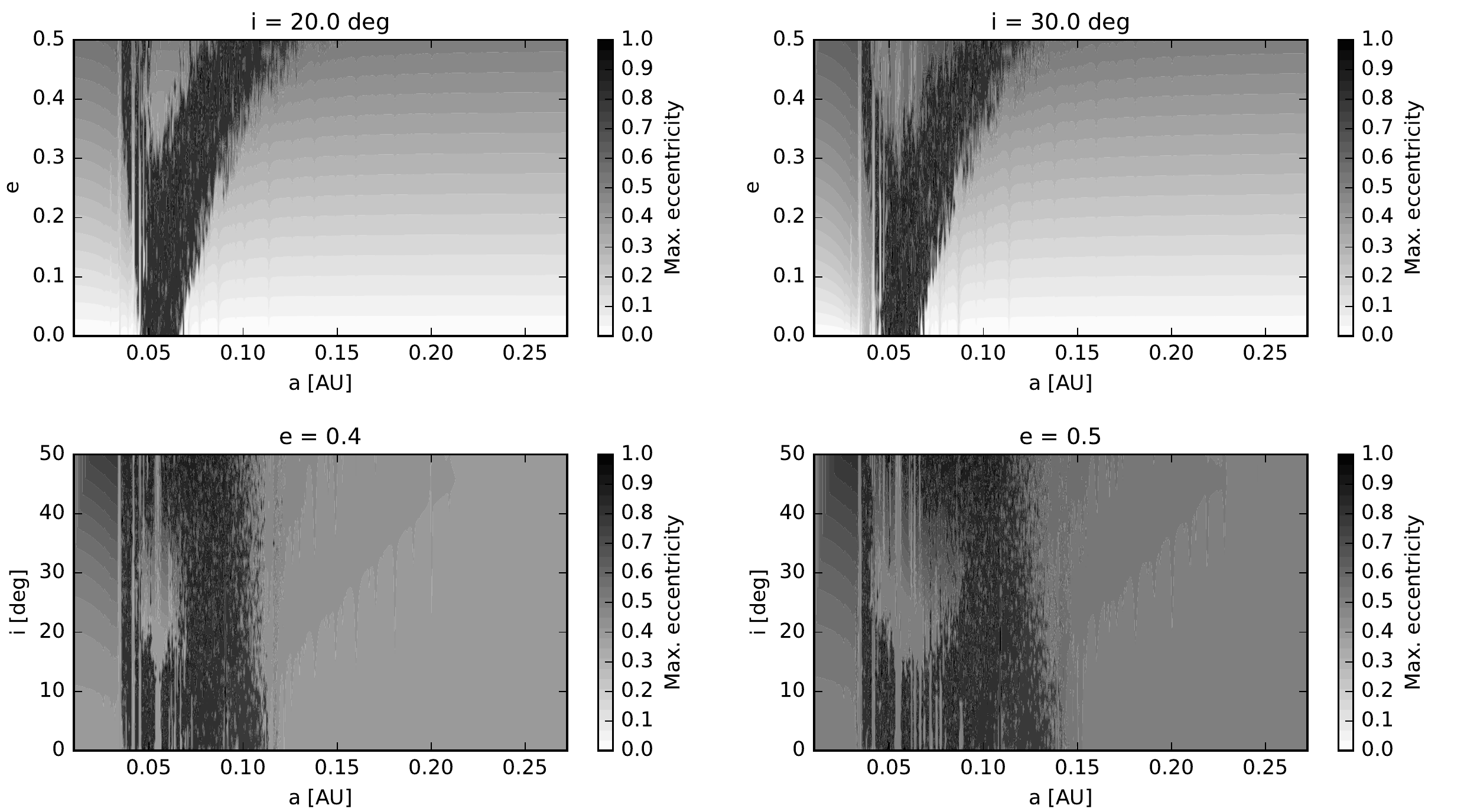}
\caption{The island of stability in WASP-118 system caused by the Kozai mechanism. \textit{Top:} Stability plot in $a-e$ plane showing the maximum eccentricity for $i = 20\degr,\ i = 30\degr$. \textit{Bottom:} Stability plot in the $a - i$ plane for $e = 0.4,\ e = 0.5$.}
\label{fig:wasp118-kozai}
\end{figure*}

In this section, we study the long-term gravitational influence of the transiting planet on another potential planet in the system. We performed numerical simulations of the system with the parent star, the known (transiting) planet and large number of massless particles representing the hypothetical lower mass (earth mass or smaller) planets.

To generate the stability maps, the method of the maximum eccentricity was used \citep{Dvorak2003}. The value of eccentricity gives us information about the probability of a close encounter between the studied body and a massive planet. An orbit with many close encounters is, of course, unstable. Other authors (e.g. \citealt{Freistetter2009}) have noted that the maximum eccentricity is good indicator of the stability of the orbit.

For each of our extrasolar planetary systems, we performed numerical integration of the orbits for $10^5$ revolutions of the known planet, giving a time span of approximately 1000 years. We again used the Bulirsch-Stoer algorithm and \textsf{MERCURY6} code. The parameter $\epsilon$ which controls the accuracy of the integration was set to $10^{-8}$. 

We adopted the parameters of giant planets from \cite{Hay2016}. We have generated 200 thousand massless particles for each studied system. The semi-major axis of our test particles ranged from 0.01 AU to 5 times the semi-major axis of the giant planet. The inner border was less than 2 times the star radius. The step for the semi-major axis was set to one hundredth of the semi-major axis of the giant planet. The upper limit of eccentricity was 0.5 and the step was 0.02. The upper limit of inclination between the orbits of the giant planet and the test particle was 50\degr, the step was 1\degr.

Figure \ref{fig:wasp92-stab} (\textit{top} and \textit{middle}) shows the stability maps in the $a - e$ plane for the WASP-92 system. We have selected maps for inclinations $i = 0\degr$ and $i = 50\degr$ as an example. The stability maps of all the studied systems for selected values of inclination ($i = 0\degr,\ i = 10\degr,\ i = 20\degr,\ i = 30\degr,\ i = 40\degr,\ i = 50\degr$) are available as supplementary material to this paper. The resulting stability maps look nearly identical (despite the differing x-axis scales) for all three systems. It is because these systems are very similar; all of them have a Jupiter-size planet on a circular orbit close to the parent star. The initial inclination of the orbit has minimal influence on the shape of the stability map in the $a - e$ plane. It plays some role in the case of orbits closer to the parent star and also for small values of initial eccentricity. Inner orbits are also stable up to inclinations of 30\degr. The highly inclined inner orbits are strongly affected by the precession caused by the known planet and therefore unstable. The presence of such a planet is also improbable from an evolutionary point of view. On the other hand, the impact of the initial eccentricity on the stability of the orbit is important. It is caused by the fact that for higher eccentric orbits there is a higher probability of a close encounter with the giant planet. The arc-like shape of the right border of the instability region is the result of close encounters between the test particle and the giant planet in the periastron of the particle orbit. There is a stable strip of coplanar orbits with high eccentricities ($e \gtrsim 0.2$) and semimajor axes nearly the same as the giant planet. This stable strip is a result of 1:1 resonance between the body in this region and known transiting planet. \cite{Freistetter2009} observed the same stable strip in their stability study of the TrES-2 system for mean anomaly $M=0\degr$ what is the same case as our analysis. Many of the bodies in the 1:1 resonance could be captured on satellite-type orbits around the giant planet. The detailed analysis of this scenario and the study of the stability of the hypothetical satellites are beyond the scope of this paper.

In figure \ref{fig:wasp92-stab} (\textit{bottom}), we present the stability map in the $a - i$ plane for the WASP-92 system for initial value of eccentricity $e = 0.0$. The supplementary materials include the stability maps in the $a - i$ plane for the studied systems for selected initial values of eccentricity ($e = 0.0,\ e = 0.1,\ e = 0.2,\ e = 0.3,\ e = 0.4,\ e = 0.5$). The width of the stable regions is generally almost independent of the inclination, mainly for lower values of eccentricity. For higher values of eccentricity, there is a wider unstable region for the lower inclinations. And for inclinations around 25 -- 30\degr and high eccentricities (0.4 and 0.5), the stable region starts to occur inside the large unstable region. This island of stability looks slightly different in each of the individual systems. These regions of stability could be caused by Kozai resonance. The Kozai mechanism affects orbits with large eccentricities and inclinations \citep{Thomas1996}. It is  strongest for the WASP-118 system (Figure \ref{fig:wasp118-kozai}).

\section{Discussion and conclusions}
\label{disc}

We redetermined the basic parameters of three exoplanets WASP-92 b, WASP-93 b and WASP-118 b. We used ground-based observations of WASP-92 and WASP-93 available at the public archive ETD. WASP-118 was observed during Campaign 8 of \textit{Kepler-K2} mission. We also determined accurate times of all transits of these planets to analyse for possible TTVs. We concluded that our values of planetary parameters for WASP-92 b and WASP-93 b are in generally consistent agreement with previous results of \cite{Hay2016}, and with \cite{Mocnik2017} for WASP-118 b.

The next aim of this paper was to discuss the  possibility of the presence of other planets in these systems. Our O-C diagrams showed no significant periodic variations. We put upper limits on the masses of potential planets in these systems which could generate a TTV signal, though the amplitude of such a signal would be lower than the observed maximum deviation of the data on our O-C diagram. We found that earth-mass planets could exist in these three systems near to MMR. High quality observations over a long period would be needed to better specify the mass limits of potential perturbing planets.

Finally, we investigated the dynamical stability of these systems in order to identify stable regions where additional planets could exist for a long time (hundreds of years). Our results for all three systems are similar, despite differing values of the semi-major axes. The differences in the shapes of the stable regions are negligible in many cases. We found that a hypothetical planet could exist relatively close to the transiting one, mainly on near-circular and near-coplanar orbits. Raising the eccentricity of the orbit causes the unstable region around the transiting planet to start to grow and for extreme cases of eccentricity ($e=0.5$) orbits with a semi-major axis up to two or three times that of the transiting planet are found to be unstable. However, the inner border of this region changes very slowly with rising eccentricity. These inner orbits are stable for low values of orbital inclination (up to 20 -- 30\degr), but the existence of another planet in this region is really questionable due to its vicinity to the parent star. But in general, the inclination has minimal influence on the shape of the regions of stability.

From a dynamic point of view, the possibility of there being other planets in these systems is high. There exists a wide range of orbits on which such planets could exist for a long time, though our analysis of times of transits excludes the presence of massive planets near to the transiting exoplanet. However additional massive planets could exist on orbits further out. The presence of planets with earth-mass (or lower mass) near the transiting ones or near the mean-motion resonance is still possible.

\section*{Acknowledgement}
We would like to thank Fran Campos, Juanjo Gonzalez, Ramon Naves and Mark Salisbury for providing the data of the studied targets. 
This paper has been supported by the grant of the Slovak Research and Development Agency with number APVV-15-0458. This article was created under project ITMS No.26220120029, based on the supporting operational Research and development program financed from the European Regional Development Fund. The research of PG was supported by internal grant VVGS-PF-2017-724 of the Faculty of Science, P. J. \v{S}af\'{a}rik University in Ko\v{s}ice. MV would like to thank the project VEGA 2/0031/18. MJ acknowledges support from the Slovak Grant Agency for Science (grant VEGA No.2/0037/18).

\bibliographystyle{mnras} 
\bibliography{document}

\section*{Supporting information}
Additional Supporting Information may be found in the on-line version of this article:\\

\noindent Figures. Stability plots in the $a - e$ and $a - i$ plane for different values of $i$ and $e$ showing the maximum eccentricity for the WASP-92, WASP-93 and WASP-118 systems.\\

\noindent Please note: Oxford University Press is not responsible for the content or functionality of any supporting 
materials supplied by the authors. Any queries (other than missing material) should be directed to the corresponding 
author for the paper.

\label{lastpage}

\onecolumn

\includepdf{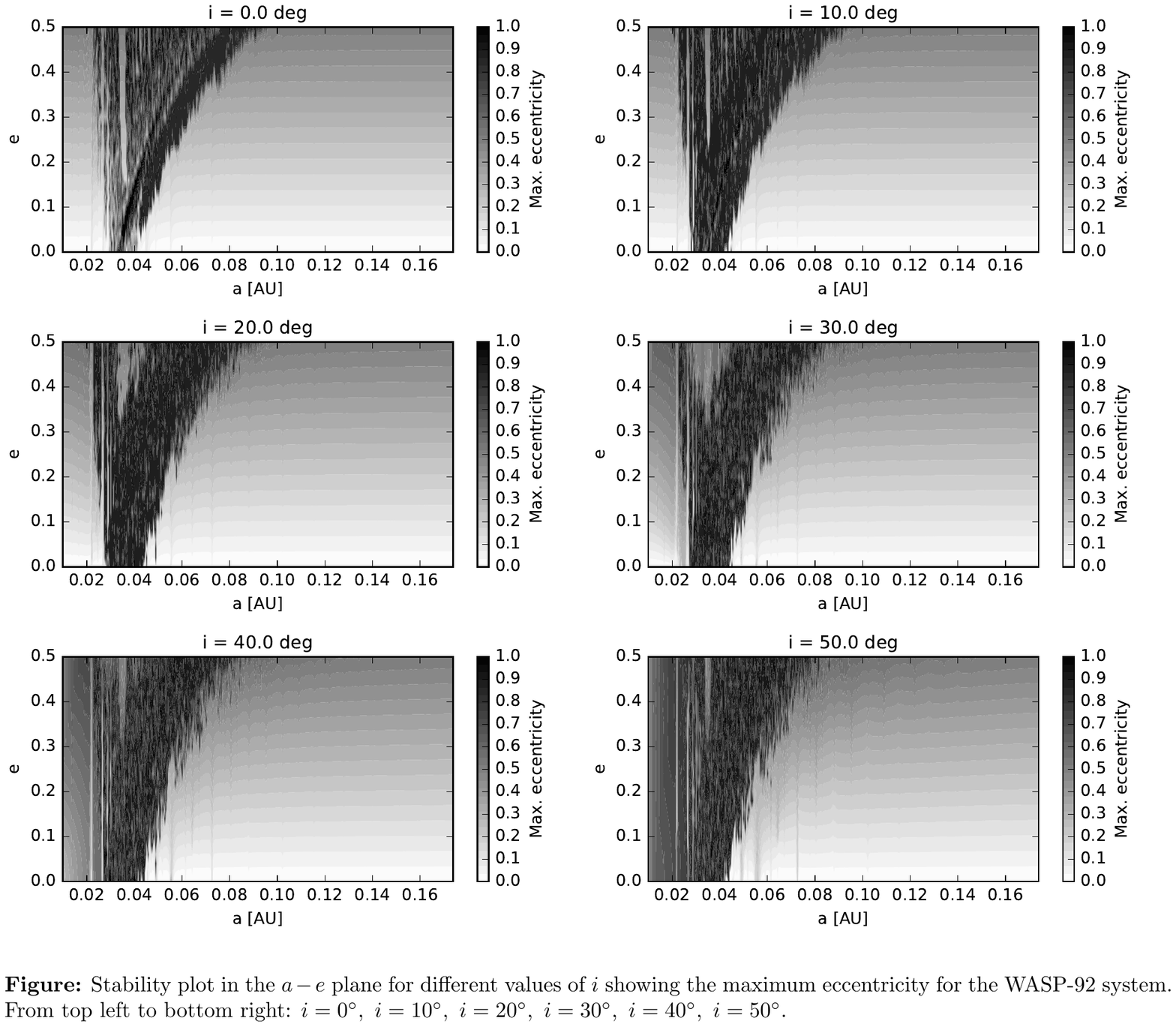}
\includepdf{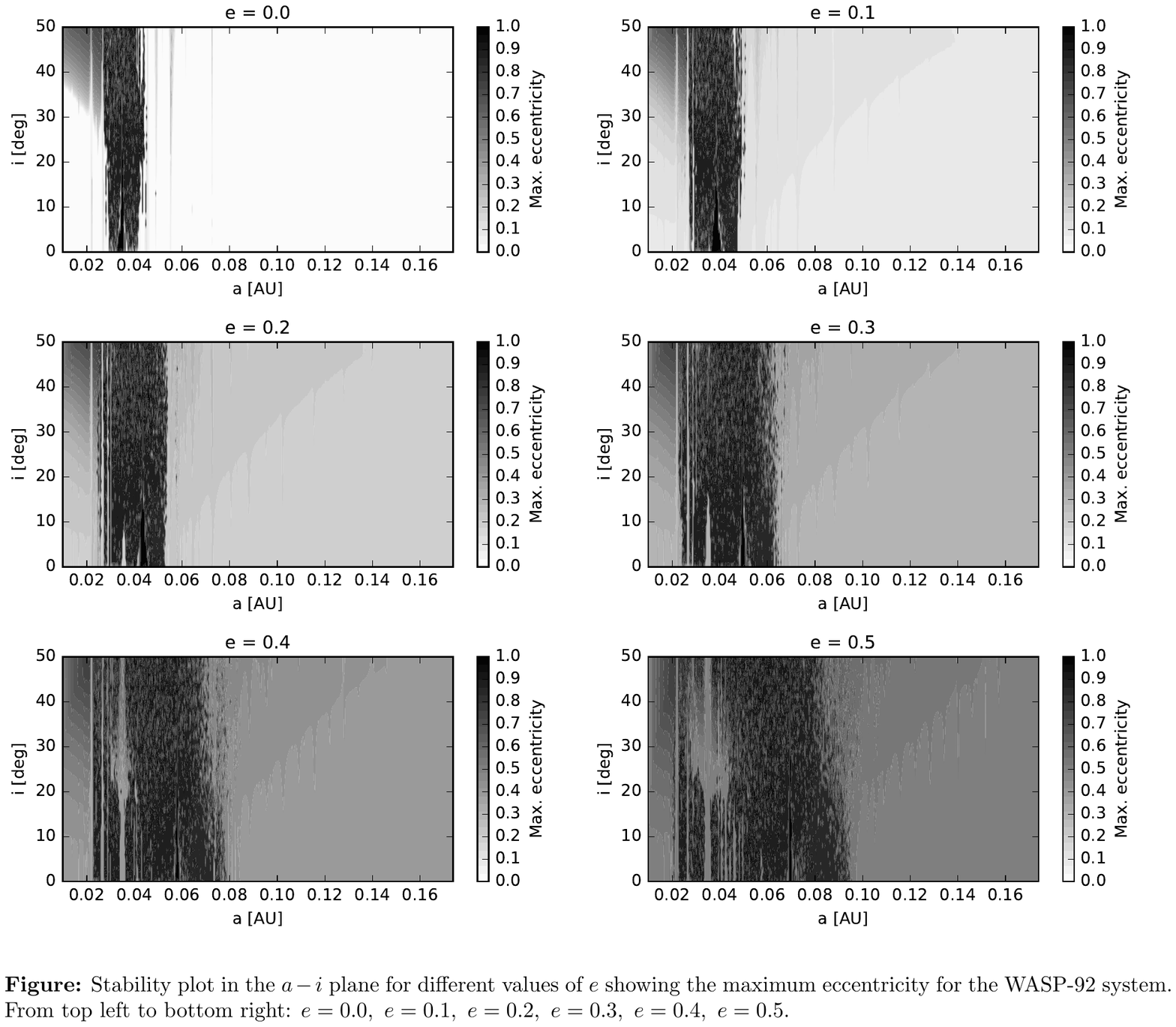}
\includepdf{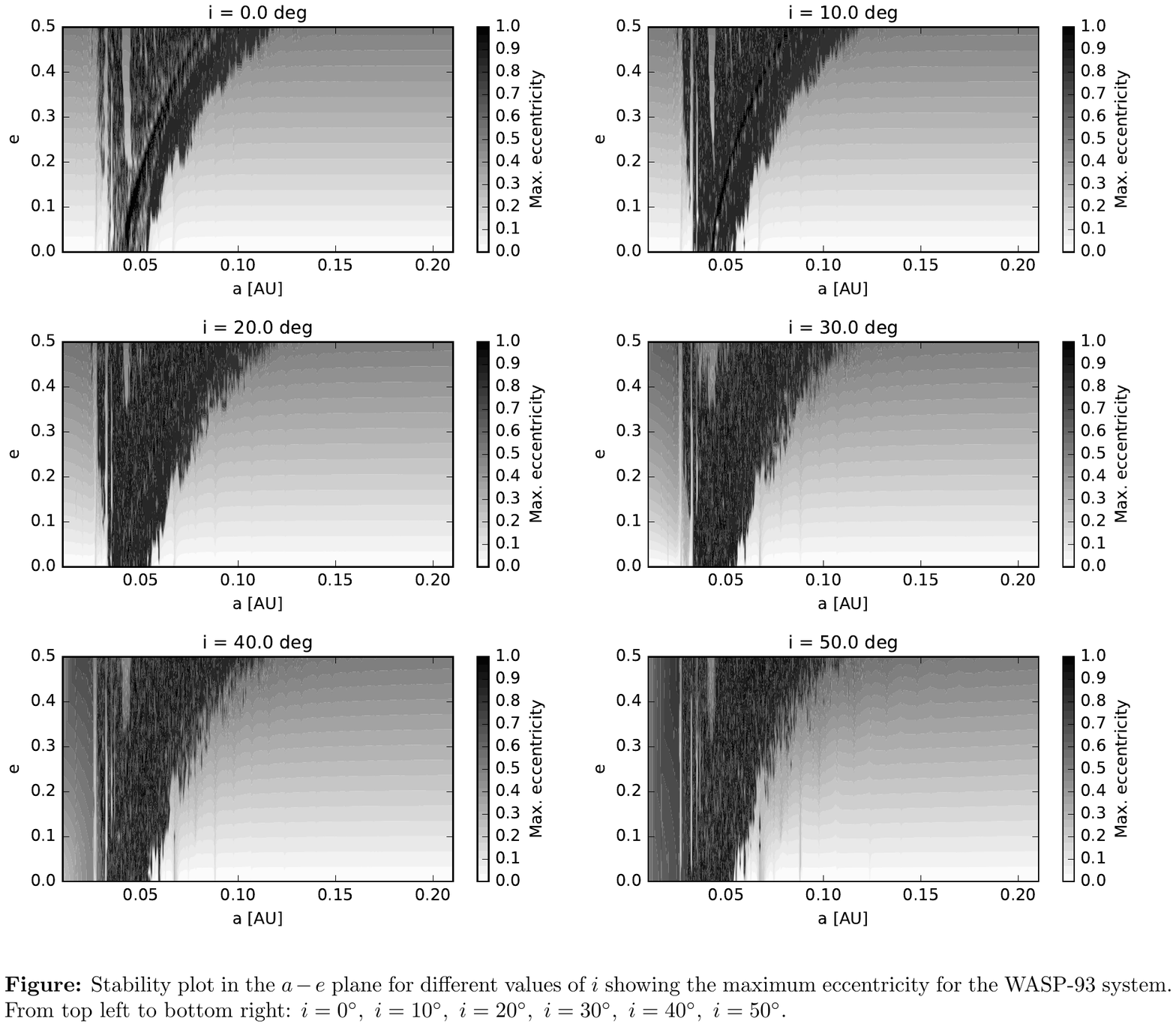}
\includepdf{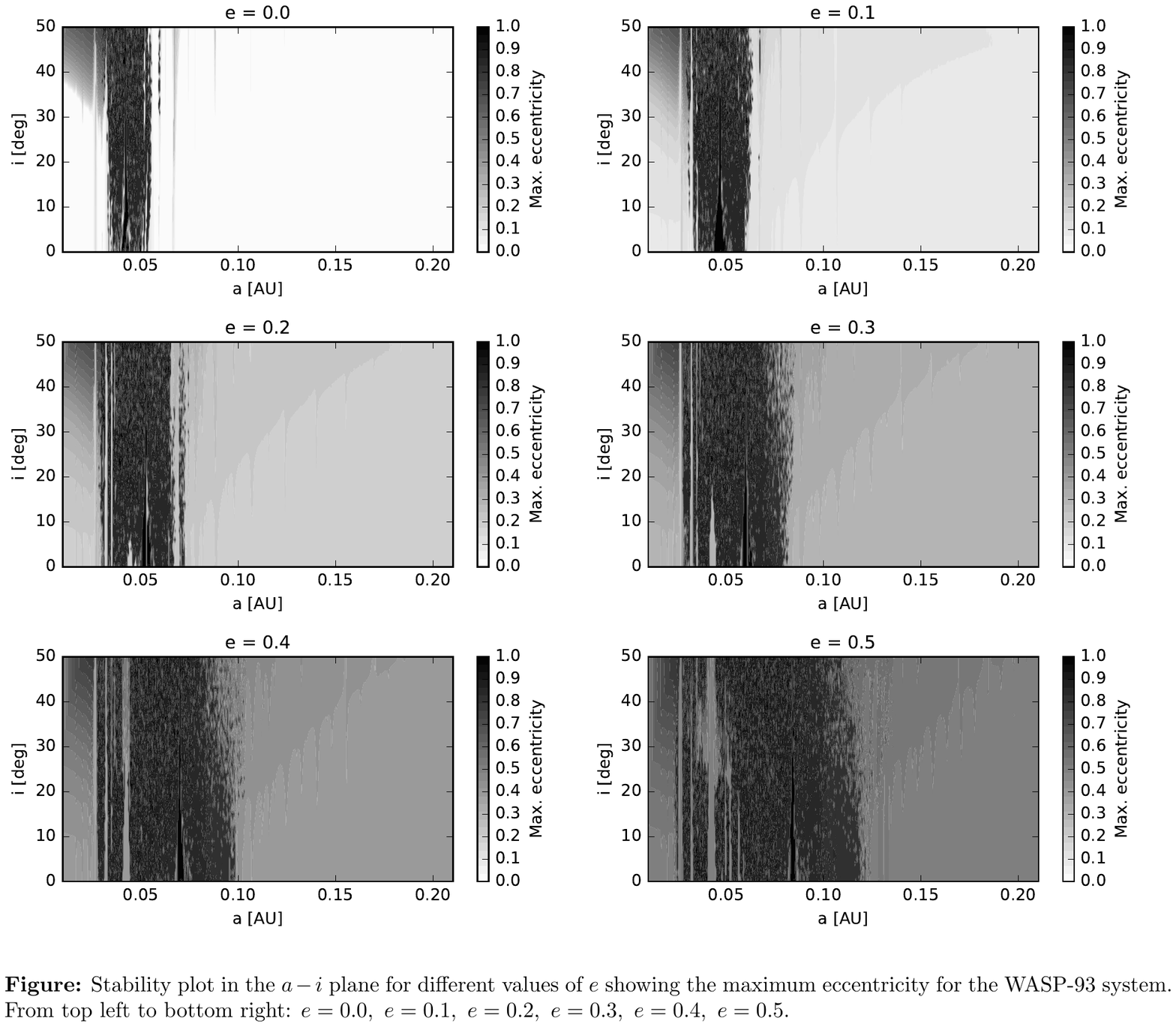}
\includepdf{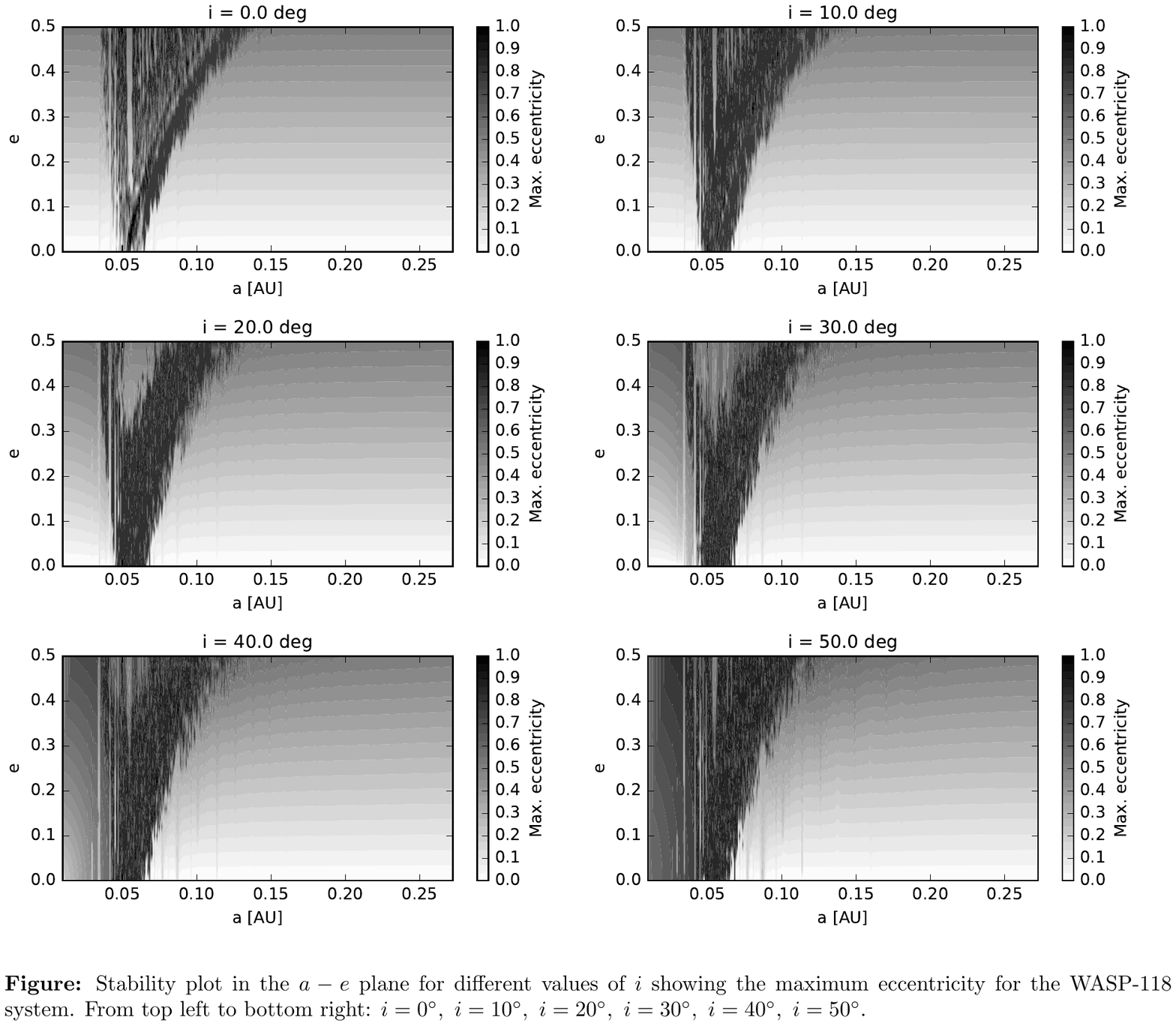}
\includepdf{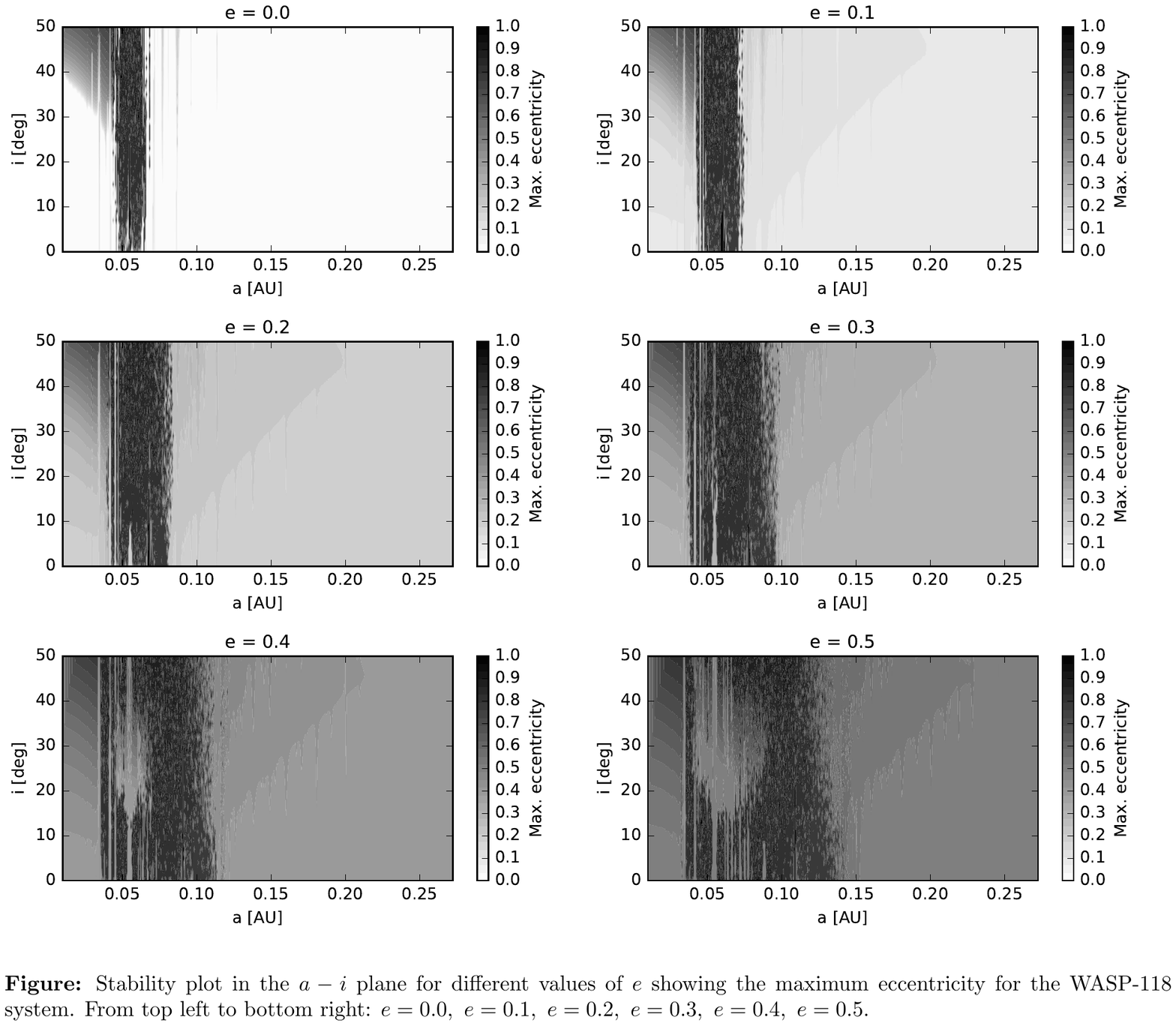}

\end{document}